
\magnification=1200
\hoffset=.0cm
\voffset=.0cm
\baselineskip=.53cm plus .53mm minus .53mm

%
%
\def\ref#1{\lbrack#1\rbrack}
%
%
%
%
\input amssym.def
\input amssym.tex
%
%
\font\teneusm=eusm10
\font\seveneusm=eusm7
\font\fiveeusm=eusm5
%
%
\font\sf=cmss10

%
%
\font\cps=cmcsc10
%
%
\newfam\eusmfam
\textfont\eusmfam=\teneusm
\scriptfont\eusmfam=\seveneusm
\scriptscriptfont\eusmfam=\fiveeusm
\def\sans#1{\hbox{\sf #1}}

\def\sh#1{\hbox{\teneusm #1}}

\def\proclaim #1. #2\par{\medbreak{\cps #1.\enspace}{\it #2}\par\medbreak}
%
%
%
%
\def\sgn{{\rm sgn}\hskip 1pt}
\def\real{{\rm Re}\hskip 1pt}

\def\ord{{\rm ord}\hskip 1pt}
\def\ker{{\rm ker}\hskip 1pt}
\def\ran{{\rm ran}\hskip 1pt}

\def\id{{\rm id}\hskip 1pt}
\def\ad{{\rm ad}\hskip 1pt}
\def\Ad{{\rm Ad}\hskip 1pt}

\def\Gau{{\rm Gau}\hskip 1pt}
\def\Conf{{\rm Conf}\hskip 1pt}
\def\Lie{{\rm Lie}\hskip 1pt}
\def\hst1{\hskip 1pt}
\def\square{\,\vbox{\hrule \hbox{\vrule height 0.25 cm
\hskip 0.25 cm \vrule height 0.25 cm}\hrule}\,}
\def\ltimes{\hbox{\hskip .04cm\vrule height 0.149cm \kern -.053cm $\times$}}
\def\rtimes{\hbox{$\times$ \kern -.053cm \vrule height 0.149cm \hskip .04cm}}
%
%
%
%

\hbox to 16.5 truecm{March 1994   \hfil DFUB 94--5}
\hbox to 16.5 truecm{Version 1  \hfil hep-th/9403036}
\vskip2cm
\centerline{\bf THE DRINFELD--SOKOLOV HOLOMORPHIC BUNDLE AND}
\centerline{\bf CLASSICAL $W$ ALGEBRAS ON RIEMANN SURFACES}
\vskip1cm
\centerline{by}
\vskip.5cm
\centerline{\bf Roberto Zucchini}
\centerline{\it Dipartimento di Fisica, Universit\`a degli Studi di Bologna}
\centerline{\it V. Irnerio 46, I-40126 Bologna, Italy}
\vskip1cm
\centerline{\bf Abstract}
Developing upon the ideas of ref. \ref{6},
it is shown how the theory of classical $W$ algebras can be formulated on
a higher genus Riemann surface in the spirit of Krichever and Novikov.
The basic geometric object is the Drinfeld--Sokolov principal bundle $L$
associated to a simple complex Lie group $G$ equipped with an $SL(2,\Bbb C)$
subgroup $S$, whose properties are studied in detail. On a multipunctured
Riemann surface, the Drinfeld--Sokolov--Krichever--Novikov spaces are
defined, as a generalization of the customary Krichever--Novikov spaces,
their properties are analyzed and standard bases are written down.
Finally, a WZWN chiral phase space based on the principal bundle $L$
with a KM type Poisson structure is introduced and, by the usual procedure
of imposing first class constraints and gauge fixing, a classical $W$
algebra is produced. The compatibility of the construction with the global
geometric data is highlighted.
\vfill\eject
\item{1.} {\bf Introduction}
\vskip.4cm
\par
During the last few years, a large body of literature has been devoted
to the study of $W$ algebras and to the understanding of their field
theoretic realizations. Originally introduced as higher
spin extensions of the Virasoro algebra, they were later shown to appear
naturally in several contexts, such as cosets of affine Lie algebras,
gauged WZWN models, Toda field theory, reductions of the KP hierarchy and,
more recently, random matrix models, string theory and $2d$ quantum gravity
(see ref. \ref{1} for a comprehensive review of the subject and extensive
referencing).

While the local properties of $W$ algebras have been the object of extensive
study, a comparatively modest effort has been made in the analysis
of their global properties so far \ref{2--5}. The present paper,
developing upon and expanding an earlier work \ref{6}, is a contribution
in such direction. The approach adopted is inspired on one hand by the seminal
work of Krichever and Novikov \ref{7}, which relies on the classical
theory of Riemann surfaces and holomorphic bundles thereupon, and on the other
by the equally seminal work of Drinfeld and Sokolov \ref{8} and
by the techniques of refs. \ref{9--13}, which use the theory of
Poisson manifolds and their reductions.
Below, I shall provide a brief account of standard results
about Toda field theory and $W$ algebras to introduce the basic concepts and
motivate the technical analysis presented in later sections (see also ref.
\ref{14} for a review).

The Toda field equations can be put in the form of a zero curvature condition
for a connection $\sh C$ satisfying a certain grading constraint.
This allows for the integrability of the Toda equations, a well established
result \ref{15}. It also hints to its geometrical nature which indeed
is describable in the language of the theory of holomorphic principal
bundles.

The basic algebraic structure of Toda equations is a simple complex
Lie group $G$ with an $SL(2,\Bbb C)$ subgroup $S$ with Lie algebras
$\goth g$ and $\goth s$, respectively. $\goth g$ is equipped with
an antiinvolution, denoted by $\dagger$, defining a compact real form
of $\goth g$ and leaving $\goth s$ invariant. $\goth s$ has standard
generators $t_{-1},~t_0,~t_{+1}$ satisfying $[t_{+1},t_{-1}]=2t_0$,
$[t_0,t_{\pm 1}]=\pm t_{\pm 1}$ and $t_d{}^\dagger=t_{-d}$.
To $t_0$, there is associated a halfinteger gradation of $\goth g$.

On a Riemann surface $\Sigma$ of higher genus with holomorphic canonical
line bundle $k$, one can define a holomorphic $G$--bundle $L^0$,
called the Drinfeld--Sokolov bundle in ref. \ref{4}, by
$$L^0{}_{ab}=k^{-t_0}{}_{ab},\eqno(1.1)$$
where $a$ and $b$ are coordinate labels.

The Toda field equations on the Riemann surface $\Sigma$ are the
zero curvature condition for the connection $\sh C=dzC+d\bar zC^*$
of $L^0$ given by
$$\eqalignno{
C=&~\partial e^\phi e^{-\phi}-\partial\ln gt_0+(1/2)t_{+1},&(1.2a)\cr
C^*=&~2e^{\ad\phi}t_{-1}g,&(1.2b)\cr}$$
where the Toda field $\phi$ is a section of $L^0$ such that $\phi^\dagger
=\phi$ and $[t_0,\phi]=0$ and $g$ is a metric of $\Sigma$ compatible with
its holomorphic structure. Explicitly,
$$\bar\partial(\partial e^\phi e^{-\phi})-\bar\partial\partial\ln gt_0
+[t_{+1},e^{\ad\phi} t_{-1}]g=0.\eqno(1.3)$$
As shown in ref. \ref{5}, this is just Hitchin's selfduality equation
for the Higgs pair $(L^0,\Omega)$, where $\Omega={1\over 2}t_{+1}$ and
the unitary connection is that of the Hermitian metric of $L^0$
given by $e^\phi g^{-t_0}$.

Let $G_-$ be the negative graded subgroup of $G$. One can show
that, on any coordinate patch, there exists a $G_-$ valued
smooth solution $\gamma$ of the equation
$$\gamma{}^{-1}\bar\partial\gamma+2e^{\ad\phi} t_{-1} g=0
\eqno(1.4)$$
such that on two overlapping coordinate domains
$$\gamma_a=L^{}_{ab}\gamma_bL^0{}_{ab}{}^{-1},
\eqno(1.5)$$
where $L$ is the holomorphic $G$--bundle defined by
$$L_{ab}=L^0{}_{ab}\exp(\partial_ak_{ab}{}^{-1}t_{-1}).
\eqno(1.6)$$
A proof of this theorem for $G=SL(n,\Bbb C)$ was given in ref. \ref{5}
but the result holds in general.
The integrability of eq. $(1.4)$ requires crucially the use of Toda
equation $(1.3)$. The solution is however non unique. $L$ was called
Drinfeld--Sokolov bundle in ref. \ref{6}. In fact, $L^0$ and $L$ are
{\it distinct} holomorphic forms of the same smooth $G$--bundle.
However, while $L^0$ has no flat form, $L$ does. Hence, $L$ admits
a {\it holomorphic} connection.

A holomorphic connection $\sh J$ of the bundle $L$ can be obtained
directly from the Toda connection $\sh C$ by a `gauge transformation'
$\gamma$ satisfying $(1.4)$. Eq. $(1.4)$ is indeed equivalent to the
vanishing of $(0,1)$ component of $\sh J=dz J+d\bar z J^*$:
$$\eqalignno{
J=&~\Ad\gamma C+\partial\gamma\gamma^{-1}
=\Ad\gamma\big(\partial e^\phi e^{-\phi}-\partial\ln gt_0+(1/2)t_{+1})
+\partial\gamma\gamma^{-1},&(1.7a)\cr
J^*=&~\Ad\gamma C^*+\bar\partial\gamma\gamma^{-1}=0.&(1.7b)\cr}$$
The zero curvature condition, equivalent to Toda equations,
now reads simply
$$\bar\partial J=0.\eqno(1.8)$$
In fact, $J$ is the $WZWN$ current and is given by
$$J=\partial h h^{-1}, \quad h=\gamma e^\phi g^{-t_0}
S\gamma'^\dagger S^{-1},\eqno(1.9)$$
where $\gamma'$ is a solution of $(1.4)$, not necessarily equal to
$\gamma$, and $S=e^{i\pi t_0}$.
Since the $G$--bundle $L$ has a large holomorphic gauge group, it is possible
to choose $\gamma$ in such a way that the current $J$ in $(1.7a)$ is of the
form
$$J=(1/2)t_{+1}-Rt_{-1}+W, \quad \ad t_{-1}W=0,\eqno(1.10)$$
where $R$ is a background holomorphic projective connection.

The above discussion shows that, in the present geometrical setting, the space
of chiral WZWN currents is to be identified with the affine space of
holomorphic connections of the holomorphic $G$--bundle $L$.
The chiral currents belonging to Toda field theory
span a subspace of such space, which is, up to holomorphic
gauge equivalence, the one defined by the constraint $(1.10)$.

The canonical Poisson structure of Toda field theory induces a Poisson
structure on the space of the Toda connection $J$ of the form $(1.10)$,
which, as a consequence, obey a classical $W$ algebra \ref{16}.
In ref. \ref{9--10}, it was shown that Toda field theory can be formulated
as a conformally invariant Hamiltonian reduction of WZWN theory and that the
classical $W$ algebra structure can be recovered in this way.
This is also the point of view adopted in this paper.

Following ref. \ref{7}, the twice punctured Riemann surface
$\Sigma\setminus\{P_-,P_+\}$ obtained from $\Sigma$ by removing two points
$P_-$ and $P_+$ in general position is considered,
generalizing the customary cylindrical setting. The appropriate WZWN phase
space consists in the affine space of meromorphic connections of the bundle
$L$ holomorphic off $P_-$ and $P_+$ equipped with a suitable Poisson structure
of Kac--Moody type.
Then, following ref. \ref{10}, the WZWN phase space is reduced by imposing
first class constraints compatible with the conformal symmetry and gauge
fixing. A classical $W$ algebra is yielded in this way.

The plan of the paper is as follows. In sect. 2, a brief account
of the basic properties of $\goth s\goth l(2,\Bbb C)$
embeddings into simple complex Lie algebras used in the sequel is given.
In sect. 3, a systematic study of the Drinfeld--Sokolov bundle is carried out.
Sect. 4 contains the basic notions of Krichever--Novikov theory and the
illustration of their generalization in the present context. Finally, in
sects. 5, 6 and 7, the theory of the WZWN phase space and its reduction
is presented, the properties of the reduced phase space are studied
and the emergence a classical $W$ algebra is shown.
\par\vskip.6cm
\item{2.} $\goth s\goth l(2,\Bbb C)$ {\bf embeddings into simple complex
Lie algebras}
\vskip.4cm
\par
In this section, I shall briefly expound the main results on the theory
of $\goth s\goth l(2,\Bbb C)$ embeddings into simple complex Lie algebras
which will be frequently relied upon in the following. A classic treatment
of the subject is provided by ref. \ref{17}.

\proclaim Remark.
In this section, $\goth g$ is a simple complex Lie algebra. $\goth s$ is an
$\goth s\goth l(2,\Bbb C)$ subalgebra of $\goth g$. $\goth c_{\goth s}$ is the
centralizer of $\goth s$ in $\goth g$.

\proclaim Theorem 2.1.
$\goth g$ is completely reducible under $\ad\goth s$.

\par
{\it Proof}. In fact, since $\goth s$ is a simple algebra, $\goth g$ is
completely reducible under $\ad\goth s$, by Weyl's theorem (th. 8, ch. III of
ref. \ref{18}). \hfill $\square$

The non triviality of $\goth c_{\goth s}$ measures the
degeneracy of the spectrum of $\goth s\goth l(2,\Bbb C)$ irreducible
representations in the reduction.

Let us denote by $\Pi$ the set of the representations of $\goth s\goth l
(2,\Bbb C)$ appearing in the reduction of $\goth g$ by $\ad \goth s$,
each counted with its multiplicity, by $j_\eta\in\Bbb Z/2$ the spin of a
representation $\eta\in\Pi$ and by $I_\eta$ the set
$\{m|m\in{\Bbb Z}/2,|m|\leq j_\eta,j_\eta-m\in{\Bbb Z}\}$. Let us further set
$j_*=\max\{j_\eta|\eta\in\Pi\}$. Since $\ad\goth s$ acts irreducibly on
$\goth s$, there is a distinguished representation $\Pi$
corresponding to $\goth s$, which will be denoted by $o$.

\proclaim Theorem 2.2.
$\goth s$ has a set of generators $t_d$, $d=-1,~0,~+1$,
satisfying the relations
$$[t_{+1},t_{-1}]=2t_0,\quad [t_0, t_{\pm 1}]=\pm t_{\pm 1}.\eqno(2.1)$$
Associated to these, there is a set $\{t_{\eta,m}|\eta\in\Pi,~m\in
I_\eta\}$ of generators of $\goth g$ such that
$$[t_d,t_{\eta,m}]=C^d_{j_\eta,m}t_{\eta,m+d},
\quad d=-1,0,+1,\eqno(2.2a)$$
$$C^{\pm 1}_{j,m}=[j(j+1)-m(m\pm 1)]^{1\over 2},\quad C^0_{j,m}=m.
\eqno(2.2b)$$
The Lie brackets of the $t_{\eta,m}$ have the following form
$$[t_{\eta,m},t_{\zeta,n}]=
\sum_{\xi\in\Pi,k\in I_\xi}F_{\eta,\zeta}{}^\xi
(j_\eta,m;j_\zeta,n|j_\xi,k)t_{\xi,k},\eqno(2.3)$$
where $(j_1,m_1;j_2,m_2|j_3,m_3)$ is a
Clebsch--Gordan coefficient and the $F_{\eta,\zeta}{}^\xi$ are constants
depending only on the $\goth s\goth l(2,\Bbb C)$ embedding $\goth s$ and
enjoying the following properties. $F_{\xi,\eta}{}^\zeta$ vanishes
unless $|j_\xi-j_\eta|\leq j_\zeta\leq j_\xi+j_\eta$ and
$j_\xi+j_\eta-j_\zeta\in\Bbb Z$. Further, for any $\xi,\eta,\zeta\in\Pi$,
$$F_{\xi,\eta}{}^\zeta=-(-1)^{j_\xi+j_\eta-j_\zeta}
F_{\eta,\xi}{}^\zeta\eqno(2.4)$$
and, for any $\xi,\eta,\zeta,\lambda\in\Pi$ and any $j\in\Bbb Z/2$, $j\geq 0$
with $|j_\xi-j_\eta|\leq j\leq j_\xi+j_\eta$ and $j_\xi+j_\eta-j\in\Bbb Z$,
$$\eqalignno{
\sum_{\mu\in\Pi}\Big\{F_{\xi,\eta}{}^\mu F_{\mu,\zeta}{}^\lambda
\delta_{j_\mu,j}+
&F_{\zeta,\xi}{}^\mu F_{\mu,\eta}{}^\lambda
\Omega(j_\xi,j_\eta,j_\zeta,j_\lambda;j,j_\mu)&\cr
+&F_{\eta,\zeta}{}^\mu F_{\mu,\xi}{}^\lambda
\Omega(j_\eta,j_\zeta,j_\xi,j_\lambda;j_\mu,j)
\Big\}=0,~~~~&(2.5)\cr}$$
where $\Omega(j_1,j_2,j_3,j_4;j_5,j_6)=(-1)^{j_3+j_5-j_4}
(2j_5+1)^{1\over 2}(2j_6+1)^{1\over 2}W(j_1,j_2,j_3,j_4;j_5,j_6)$
and $W(j_1,j_2,j_3,j_4;j_5,j_6)$ is a Racah--Wigner function $\ref{19}$.
Finally, one has
$$t_{o,\pm 1}=\mp 2^{-{1\over 2}}t_{\pm 1},\quad t_{o,0}=t_0,\eqno(2.6)$$
$$F_{o,\eta}{}^\zeta=F_{\eta,o}{}^\zeta=-\delta_{\eta,\zeta}
[j_\eta(j_\eta+1)]^{1\over 2}\eqno(2.7)$$

\par
{\it Proof}. $(2.1)$ and $(2.2)$ are standard results from the representation
theory of $\goth s\goth l(2,\Bbb C)$ \ref{18--19}.
Let $[t_{\eta,m},t_{\zeta,n}]=
\sum_{\xi\in\Pi,k\in I_\xi}F_{\eta,m;\zeta,n}{}^{\xi,k}t_{\xi,k}$,
where the $F_{\eta,m;\zeta,n}{}^{\xi,k}$ are structure constants.
{}From the Jacobi identity for the triple of generators $t_d,t_{\eta,m},
t_{\zeta,n}$, one gets
$$C^{-d}_{j_\xi,k}F_{\eta,m;\zeta,n}{}^{\xi,k-d}
-C^d_{j_\eta,m}F_{\eta,m+d;\zeta,n}{}^{\xi,k}
-C^d_{j_\zeta,n}F_{\eta,m;\zeta,n+d}{}^{\xi,k}=0,\quad d=-1,0,+1.\eqno(2.8)$$
For fixed $\xi,~\eta,~\zeta\in\Pi$, such relations have the same form as the
recurrence relation of the Clebsch--Gordan coefficients. This yields
$(2.3)$. $(2.4)$ and $(2.5)$ follow from the antisymmetry and the Jacobi
identity of the Lie brackets and from well-known properties of the
Clebsch--Gordan coefficients: $(j_2,m_2;j_1,m_1|j_3,m_3)=
(-1)^{j_1+j_2-j_3}(j_1,m_1;j_2,m_2|j_3,m_3)$;
$(j_1,m_1;j_2,m_2|j_{12},m_1+m_2)(j_{12},m_1+m_2;j_3,m_3|j_4,m_4)
=\sum_{j_{23}}(j_2,m_2;j_3,m_3|j_{23},m_2+m_3)(j_1,m_1;j_{23},m_2+m_3|j_4,m_4)
(2j_{12}+1)^{1\over 2}(2j_{23}+1)^{1\over 2}W(j_1,j_2,j_4,j_3;j_{12},j_{23})$
\ref{19}. $(2.6)$ follows from comparing $(2.1)$ and $(2.2)$. $(2.7)$ follows
$(2.6)$ and from comparing $(2.3)$ and $(2.2)$. \hfill $\square$

Denote by $(\cdot,\cdot)$ the Cartan--Killing form of $\goth g$.

\proclaim Theorem 2.3. One has
$$(t_{+1},t_{-1})=2(t_0,t_0).\eqno(2.9)$$
For each representation $\eta\in\Pi$, there is a
conjugate representation $\bar\eta$ such that $j_\eta=j_{\bar\eta}$.
Further $\bar{\bar\eta}=\eta$ and
$\bar\eta=\eta$ if and only if $j_\eta\in{\Bbb Z}$. Moreover, for $\eta,
{}~\zeta\in\Pi$, $m\in I_\eta$ and $n\in I_\zeta$
$$\big(t_{\eta,m},t_{\zeta,n}\big)=N_\eta\delta_{\eta,\bar\zeta}
(-1)^{j_\eta-m}\delta_{m,-n},\eqno(2.10a)$$
where $N_\eta$ is a normalization constant such that
$$N_{\bar\eta}=(-1)^{2j_\eta}N_\eta.\eqno(2.10b)$$
In particular, one has $o=\bar o$ and
$$N_o=-(t_0,t_0).\eqno(2.11)$$

\par
{\it Proof}. $(2.9)$ follows easily from $(2.1)$ and the $\ad$ invariance of
the Cartan--Killing form.
For any homogeneous polynomial $P$ of degree $p$ in
$\ad t_d$, $d=-1,~0,~+1$, one has $(Px,y)+(-1)^p(x,Py)=0$
for all $x,~y\in\goth g$. Choosing $P=\ad t_0,~
{1\over 2}(\ad t_{-1}\ad t_{+1}+\ad t_{+1}\ad t_{-1})+(\ad t_0)^2$
and $x=t_{\eta,m}$ and $y=t_{\zeta,n}$, one finds
$$(t_{\eta,m},t_{\zeta,n})=N_{\eta,\zeta,m}\delta_{j_\eta,j_\zeta}
\delta_{m,-n}.\eqno(2.12)$$
Choosing for $P=\ad t_{+1}$ and $x=t_{\eta,m-1}$ and $y=t_{\zeta,-m}$
and using $(2.12)$, one finds further
$$N_{\eta,\zeta,m}=N_{\eta,\zeta}(-1)^{j_\eta-m}.\eqno(2.13)$$
The non singularity of the Cartan--Killing form implies that
the matrix $N_{\eta,\zeta}$ is non singular. From $(2.12)$ and
$(2.13)$, it follows that
$$N_{\eta,\zeta}=(-1)^{2j_\eta}N_{\zeta,\eta},\eqno(2.14)$$
where $j_\eta=j_\zeta$.
Hence, the matrix $N_{\eta,\zeta}$ is either symmetric or antisymmetric.
By a congruence, it can be put in the form of either diagonal matrix
with non zero diagonal entries or a direct sum of matrices of the form
$i\sigma_2$ with a non zero coefficient, respectively, where $\sigma_2$
is a Pauli matrix. In both cases, for each $\eta\in\Pi$ there
exists a unique $\bar\eta\in\Pi$ such that $N_{\eta,\bar\eta}\not=0$.
{}From here, $(2.10)$ follows easily. The remaining statements are
obvious. \hfill $\square$

Associated to the $\goth s\goth l(2,\Bbb C)$ subalgebra $\goth s$
of $\goth g$, there is a halfinteger grading of $\goth g$. For any
$m\in\Bbb Z/2$, one sets $\goth g_m=\bigoplus_{\eta\in\Pi,m\in I_\eta}
\Bbb C\hst1 t_{\eta,m}$. This is just the eigenspace of $\ad t_0$ with
eigenvalue $m$. Note that, $\goth g_m=0$ for $|m|>j_*$. One also introduces
the subspaces $\goth g_{<m}=\bigoplus_{k<m}\goth g_k$, $\goth g_{\not=m}
=\bigoplus_{k\not=m}\goth g_k$, etc..

It is readily seen that $\goth g_0$ is a subalgebra of $\goth g$.
For any $m\in\Bbb Z/2$ with $m>0$, $\goth g_{<-m}$ and $\goth g_{>m}$
are nilpotent subalgebras of $\goth g$. It can be verified that
$\goth c_{\goth s}=\bigoplus_{\eta\in\Pi,j_\eta=0}\Bbb C\hst1 t_{\eta,0}$.
In particular, $\goth c_{\goth s}$ is a subalgebra
of $\goth g_0$. One also has the identity $\ker\ad t_{\pm}
=\bigoplus_{\eta\in\Pi}\Bbb C\hst1 t_{\eta,\pm j_\eta}$.

For principal $\goth s\goth l(2,\Bbb C)$ embeddings, $\goth g_0$ is a Cartan
subalgebra of $\goth g$. Further, $\goth c_{\goth s}$ is trivial,
$\Pi$ contains only integer spin representations of strictly
positive spin with unit multiplicity. This is non longer true for non
principal $\goth s\goth l(2,\Bbb C)$ embeddings.
\par\vskip.6cm
\item{3.} {\bf The Drinfeld--Sokolov holomorphic $G$ bundle and its
properties}
\vskip.4cm
\par
This section is dedicated to the study of the main properties of the
Drinfeld--Sokolov (DS) bundle, which is the basic geometric object entering
in the construction of classical $W$ algebras illustrated in sects. 5, 6
and 7. The analysis developed below envisages only the local properties of
$SL(2,\Bbb C)$ embeddings into simple complex Lie groups and, thus,
is amenable by the Lie algebraic methods developed in sect. 2.

\proclaim Remark.
Throughout this section, the following assumptions are made.
$\Sigma$ is a compact connected Riemann surface without boundary of genus
$\ell\geq 2$. $k^{\otimes{1\over 2}}$ is a fixed theta characteristic.
$h$ is a fixed element of $\Bbb Z/2$. $G$ is a connected simple
complex Lie group. $S$ is an $SL(2,\Bbb C)$ subgroup of $G$.

Recall that $k^{\otimes{1\over 2}\otimes 2}=k$, where $k$ is the holomorphic
canonical $1$--cocycle of $\Sigma$ defined by $k_{ab}=\partial_az_b$.

I denote by $z$ the generic holomorphic coordinate of $\Sigma$ and by
$\partial$ the operator $\partial/\partial z$. I further use lower Latin
indices $a$, $b$, $c$, ... as labels for different coordinates. Further,
$k^{\otimes h}$ is short for $k^{\otimes{1\over 2}\otimes 2h}$. For any
holomorphic $1$--cocycle $K$ on $\Sigma$ representing some holomorphic bundle
on $\Sigma$ and any non empty open set $U$ of $\Sigma$, I denote by
$\Gamma(U,{\cal O}(K))$ and $\Gamma(U,{\cal M}(K))$ the spaces of holomorphic
and meromorphic sections of $K$ on $U$, respectively. Finally, I denote by
$\exp$ the exponential map of $G$ and by $C_S$ the centralizer of $S$ in $G$.

\proclaim 1.
The DS holomorphic $G$--bundle $L$

\proclaim Definition 3.1.
Let $t_{-1},t_0,t_{+1}$ be the standard generators
of $\goth s$. For any two overlapping coordinate domains, one sets
$$L_{ab}=\exp(-2\ln k^{\otimes{1\over 2}}{}_{ab}t_0)
\exp(\partial_ak_{ab}{}^{-1}t_{-1}),\eqno(3.1)$$
where $\exp$ is the exponential map of $G$.

\proclaim Theorem 3.1.
$L=\{L_{ab}\}$ is a holomorphic $G$--valued
$1$--cocycle on $\Sigma$. It thus defines a holomorphic $G$--bundle
canonically associated to the pair $(G,S)$.

\par
{\it Proof}. One has $\exp(4\pi it_0)=1$.
Further, $k^{\otimes{1\over 2}}$ is a holomorphic $1$--cocycle
on $\Sigma$. From these facts, it is easily checked that
$\{\exp(-2\ln k^{\otimes{1\over 2}}{}_{ab} t_0)\}$
is a holomorphic $G$--valued $1$--cocycle on $\Sigma$. Using $(2.1)$, it is
then straightforward to verify that $\{L_{ab}\}$, also, is a holomorphic
$G$--valued $1$--cocycle on $\Sigma$. \hfill $\square$

$L$ will be called the DS bundle \ref{4,6,20}.

In application to classical $W$ algebras, the relevant $1$--cocycles are of
the form $k^{\otimes h}\otimes \Ad L$, where $h\in \Bbb Z/2$. Below, I shall
carry out a systematic study of them.

\proclaim 2.
Generalities on $\Gamma(\Sigma,{\cal M}(k^{\otimes h}\otimes\Ad L))$

Let $\Phi\in\Gamma(\Sigma,{\cal M}(k^{\otimes h}\otimes \Ad L))$.
$\Phi$ can be expanded in the basis $\{t_{\eta,m}|\eta\in\Pi, m\in I_\eta\}$ of
$\goth g$ canonically associated to its $\goth s\goth l(2,\Bbb C)$ subalgebra
$\goth s$, obtaining
$$\Phi_a=\sum_{\eta\in\Pi, m\in I_\eta}\Phi_{\eta,m\hst1 a}t_{\eta,m},
\eqno(3.2)$$
where the $\Phi_{\eta,m\hst1 a}$ are certain meromorphic functions.

\proclaim Theorem 3.2.
For any $\Phi\in\Gamma(\Sigma,{\cal M}(k^{\otimes h}
\otimes \Ad L))$, one has
$$\Phi_{\eta,m\hst1 a}=k^{\otimes h}{}_{ab}\sum_{n\in I_\eta}L^{(\eta)}{}_{ab
\hst1 m}{}^n\Phi_{\eta,n\hst1 b},\eqno(3.3)$$
where $L^{(\eta)}=\{L^{(\eta)}{}_{ab}\}$ is the holomorphic
$SL(2j_\eta+1,\Bbb C)$--valued $1$--cocycle
$$\eqalignno{L^{(\eta)}{}_{ab\hst1 m}{}^n=&~
{1\over (n-m)!}\bigg(\prod_{r\in\Bbb N,1\leq r\leq
n-m}C^{+1}_{j_\eta,n-r}\bigg)
k^{\otimes-m}{}_{ab}(\partial_ak_{ab}{}^{-1})^{n-m},\quad m,n\in I_\eta,~
m\leq n,&\cr
&&(3.4a)\cr
L^{(\eta)}{}_{ab\hst1 m}{}^n=&~0\vphantom{1\over (n-m)!},\quad m,n\in I_\eta,~
m>n,&(3.4b)\cr}$$
where $C^d_{j,k}$ is given by $(2.2b)$.

\par
{\it Proof}. This follows easily
from substituting the expansion $(3.2)$ into the
relation $\Phi_a=k^{\otimes h}{}_{ab}\Ad L_{ab}\Phi_b$ and then using $(2.2)$.
The calculation is straightforward. \hfill $\square$

The following technical theorem will be of crucial importance in the following
treatment.

Recall that a projective connection $R$ is a holomorphic $0$--cochain
$\{R_a\}$ such that $R_a=k_{ab}{}^2(R_b-\{z_a,z_b\})$, where $\{z_a,z_b\}
=\partial_b{}^2\ln\partial_bz_a-{1\over 2}(\partial_b\ln\partial_bz_a)^2$ is
the Schwarzian.

\proclaim Theorem 3.3.
Let $R$  be a holomorphic projective connection. Let
$\eta\in\Pi$ and $\mu\in I_\eta$ with either $\mu<h$ or $\mu\geq j_\eta+2h$.
Let $\phi\in\Gamma(\Sigma,{\cal M}(k^{\otimes h-\mu}))$. Then, there exists a
unique element $\Phi\in\Gamma(\Sigma,{\cal M}(k^{\otimes h}\otimes \Ad L))$
such that
$$\eqalignno{\Phi_{\eta,m}=&~\phi\delta_{\mu,m}
\vphantom{C^{+1}_{j_\eta,m}\over g_{\mu-h,m-h}},\quad m\in I_\eta,~m\geq\mu,
&(3.5a)\cr
\Phi_{\eta,m}=&~{C^{+1}_{j_\eta,m}\over g_{\mu-h,m-h}}
\Big(\partial\Phi_{\eta,m+1}+C^{+1}_{j_\eta,m+1}R\Phi_{\eta,m+2}\Big),
\quad m\in I_\eta,~m<\mu,&(3.5b)\cr
\Phi_{\zeta,n}=&~0\vphantom{C^{+1}_{j_\eta,m}\over g_{\mu-h,m-h}},
\quad \zeta\in\Pi,~\zeta\not=\eta,~n\in I_\zeta,
&(3.5c)\cr}$$
where $g_{x,y}={1\over 2}(x(x+1)-y(y+1))$. $\Phi$ depends linearly on $\phi$.
Moreover, if $\phi\in\Gamma(\Sigma,{\cal O}(k^{\otimes h-\mu}))$, then $\Phi
\in\Gamma(\Sigma,{\cal O}(k^{\otimes h}\otimes \Ad L))$.

\par
{\it Proof}. $g_{\mu-h,m-h}$ vanishes for $m=\mu, -\mu+2h-1$. So, $g_{\mu-h,
m-h}$ will vanish for some $m\in I_\eta$ with $m<\mu$ if $-j_\eta\leq-\mu+2h-1<
\mu$, {\it i. e.} $-{1\over 2}+h<\mu\leq j_\eta+2h-1$. The latter relation,
however, cannot be fulfilled by the assumptions made on $\mu$. Hence,
$g_{\mu-h,
m-h}\not=0$ in the range of $m$ values indicated. Further, $(3.5b)$ provides
a recurrence relation for the components $\Phi_{\eta,m}$ with $m\leq\mu$
with $(3.5a)$ as initial condition. This allows the unique determination of all
$\Phi_{\eta,m}$ in terms of $\phi$, $R$ and their derivatives. It is also
apparent that $\Phi_{\eta,m}$ is meromorphic. To complete the proof, one has
only to show that the $\Phi_{\eta,m}$, as determined by $(3.5b)$, glue
according to $(3.3)$ with $L^{(\eta)}$ given by $(3.4)$. The verification
is trivial for $m\geq\mu$. To show that the statement is true also for $m<\mu$,
one proceeds by induction. Suppose that one has been able to show that the
$\Phi_{\eta,n}$ glue according to $(3.3)$ for $n\in I_\eta$ with $m\leq n$,
where $m\in I_\eta$ with $-j_\eta<m\leq\mu$. By using this information,
let us show that $\Phi_{\eta,m-1}$, also, glues according to $(3.3)$. Now,
since $m-1\in I_\eta$ and $m-1<\mu$, one can use $(3.5b)$. Using the inductive
hypothesis and $(3.5a)$ one computes
$$\eqalignno{
&\Phi_{\eta,m-1\hst1 a}={C^{+1}_{j_\eta,m-1}\over g_{\mu-h,m-1-h}}
k^{\otimes 1+h}{}_{ab}\bigg\{-h\partial_ak_{ab}{}^{-1}\cdot\sum_{n\in I_\eta,
m\leq n\leq\mu}L^{(\eta)}{}_{ab\hst1 m}{}^n\Phi_{\eta,n\hst1 b}
&\cr&
+\sum_{n\in I_\eta,m\leq n\leq\mu}\partial_bL^{(\eta)}{}_{ab\hst1 m}{}^n
\Phi_{\eta,n\hst1 b}
+\sum_{n\in I_\eta,m-1\leq n\leq\mu-1}\bigg({g_{\mu-h,n-h}\over C^{+1}_{j_\eta,
n}}\bigg)L^{(\eta)}{}_{ab\hst1 m}{}^{n+1}\Phi_{\eta,n\hst1 b}
&\cr&
-R_b\cdot\sum_{n\in
I_\eta,m+1\leq n\leq\mu}C^{+1}_{j_\eta,n-1}L^{(\eta)}{}_{ab\hst1 m}{}^{n-1}
\Phi_{\eta,n\hst1 b}\vphantom{g_{\mu-h,n-h}\over C^{+1}_{j_\eta,n}}
&\cr&
+C^{+1}_{j_\eta,m}k_{ab}(R_b-\{z_a,z_b\})\cdot
\sum_{n\in I_\eta,m+1\leq n\leq\mu}
L^{(\eta)}{}_{ab\hst1 m+1}{}^{n}\Phi_{\eta,n\hst1 b}\vphantom{g_{\mu-h,n-h}
\over C^{+1}_{j_\eta,n}}\bigg\}.&(3.6)\cr}$$
Let us compute $\partial_bL^{(\eta)}{}_{ab\hst1 m}{}^n$. To this end, one uses
$(3.4)$ and the identities $k_{ab}{}^{-1}\partial_a{}^2k_{ab}{}^{-1}
-{1\over 2}(\partial_ak_{ab}{}^{-1})^2=\{z_a,z_b\}$ and $C^{+1}_{j,i-1}{}^2
-C^{+1}_{j,l}{}^2=(l+i)(l-i+1)$, which follow easily from the definition
of the Schwarzian and from $(2.2b)$, respectively. One finds
$$\eqalignno{\partial_bL^{(\eta)}{}_{ab\hst1 i}{}^l
=&~-(C^{+1}_{j_\eta,l}/2)L^{(\eta)}{}_{ab\hst1 i}{}^{l+1}
+(C^{+1}_{j_\eta,i-1}/2)k_{ab}{}^{-1}L^{(\eta)}{}_{ab\hst1 i-1}{}^l&\cr
&+C^{+1}_{j_\eta,i}k_{ab}\{z_a,z_b\}L^{(\eta)}{}_{ab\hst1 i+1}{}^l,
\quad i,l\in I_\eta,~i\leq l,&(3.7)\cr}$$
where the first term vanishes for $n=j_\eta$.
{}From $(3.4)$, it is straightforward to verify also the identities
$$C^{+1}_{j_\eta,i}k_{ab}L^{(\eta)}{}_{ab\hst1 i+1}{}^l=
C^{+1}_{j_\eta,l-1}L^{(\eta)}{}_{ab\hst1 i}{}^{l-1},
\quad i,l\in I_\eta,~i+1\leq l,\eqno(3.8)$$
$$C^{+1}_{j_\eta,i-1}\partial_ak_{ab}{}^{-1}L^{(\eta)}{}_{ab\hst1 i}{}^l=
(l-i+1)k_{ab}{}^{-1}L^{(\eta)}{}_{ab\hst1 i-1}{}^l
\quad i,l\in I_\eta,~i\leq l.\eqno(3.9)$$
Using $(3.7)$--$(3.9)$ in $(3.6)$ and performing some simplifications, one
obtains
$$\eqalignno{
&\Phi_{\eta,m-1\hst1 a}={C^{+1}_{j_\eta,m-1}\over g_{\mu-h,m-1-h}}
k^{\otimes 1+h}{}_{ab}\bigg\{\sum_{n\in I_\eta,m\leq n\leq\mu}
\bigg[-{C^{+1}_{j_\eta,n}\over 2}L^{(\eta)}{}_{ab\hst1 m}{}^{n+1}
&\cr&
+\bigg(-h{n-m+1\over C^{+1}_{j_\eta,m-1}}+{C^{+1}_{j_\eta,m-1}\over 2}
\bigg)k_{ab}{}^{-1}L^{(\eta)}{}_{ab\hst1 m-1}{}^n
\bigg]\Phi_{\eta,n\hst1 b}
&\cr&
+\sum_{n\in I_\eta,m-1\leq n\leq\mu-1}
\bigg({g_{\mu-h,n-h}\over C^{+1}_{j_\eta,n}}\bigg)
L^{(\eta)}{}_{ab\hst1 m}{}^{n+1}\Phi_{\eta,n\hst1 b}\bigg\}.&(3.10)\cr}$$
Employing $(3.8)$ to express $L^{(\eta)}{}_{ab\hst1 m}{}^{n+1}$ in terms
of $L^{(\eta)}{}_{ab\hst1 m-1}{}^n$ one gets, after a little algebra,
$$\Phi_{\eta,m-1\hst1 a}=k^{\otimes h}{}_{ab}
\sum_{n\in I_\eta,m-1\leq n\leq\mu}L^{(\eta)}{}_{ab\hst1 m-1}{}^n
\Phi_{\eta,n\hst1 b}. \eqno(3.11)$$
By induction the proof is completed. The remaining statements are
obvious. \hfill $\square$

\proclaim Definition 3.2.
For $R$, $\eta$, $\mu$ and $\phi$ as in th. 3.3, let $F_{h,\eta,\mu}(\phi|R)$
be the unique element of $\Phi\in\Gamma(\Sigma,{\cal M}(k^{\otimes h}\otimes
\Ad L))$ satisfying $(3.5)$.

By explicit calculation, one finds
$$\eqalignno{F_{h,\eta,\mu}(\phi|R)_{\eta,m}=&~\phi\delta_{\mu,m},
\quad m\in I_{\eta},~m\geq\mu,&(3.12a)\cr
F_{h,\eta,\mu}(\phi|R)_{\eta,m}=&~N_{h,\eta,\mu,m}
D_{h,\mu,\mu-m}(R)\phi\quad m\in I_{\eta},~m<\mu,&(3.12b)\cr}$$
where
$$N_{h,\eta,\mu,m}=\prod_{n\in I_\eta,m\leq n\leq\mu-1}
{C^{+1}_{j_\eta,n}\over g_{\mu-h,n-h}},\eqno(3.12c)$$
$$\eqalignno{D_{h,\mu,1}(R)=&~\partial,&\cr
D_{h,\mu,2}(R)=&~\partial^2+g_{\mu-h,\mu-h-1}R,&\cr
D_{h,\mu,3}(R)=&~\partial^3+(g_{\mu-h,\mu-h-1}+g_{\mu-h,\mu-h-2})R
\partial+g_{\mu-h,\mu-h-1}(\partial R),&\cr
D_{h,\mu,4}(R)=&~\partial^4+(g_{\mu-h,\mu-h-1}+g_{\mu-h,\mu-h-2}
+g_{\mu-h,\mu-h-3})R\partial^2
&\cr+&(2g_{\mu-h,\mu-h-1}+g_{\mu-h,\mu-h-2})(\partial R)\partial
+g_{\mu-h,\mu-h-1}\big((\partial^2 R)+g_{\mu-h,\mu-h-3}R^2\big)
&\cr&~{\rm etc.}.&(3.12d)\cr}$$
These operators provide non trivial generalizations of the standard
Bol operators \ref{21}.

{}From the above, one deduces the following theorem.

For any $w\in\Bbb Z/2$ such that $w\geq 0$, let $J(k^{\otimes-w})$ denote the
$2w$--th jet extension of $k^{\otimes-w}$, {\it i. e.} the holomorphic
$SL(2w+1,\Bbb C)$--valued $1$--cocycle defined by $\partial_a{}^m\phi_a=
\sum_{n=0}^{2w}J(k^{\otimes-w})_{ab\hst1 m}{}^n\partial_b{}^n\phi_b$,
$m=0,~1,\cdots,~2w$ for any $\phi\in\Gamma(\Sigma,{\cal M}(k^{\otimes-w}))$.

\proclaim Theorem 3.4.
One has the direct sum decomposition
$$\Ad L\cong\bigoplus_{\eta\in\Pi}L^{(\eta)}.\eqno(3.13)$$
Further, for any $\eta\in\Pi$, one has the holomorphic equivalence
$$L^{(\eta)}\cong J(k^{\otimes-j_\eta}).\eqno(3.14)$$

\par
{\it Proof}. $(3.13)$ follows from $(3.3)$ and $(3.4)$ directly.
Choose a holomorphic projective connection $R$. From $(3.5)$,
it is easily verified that, for any $\eta\in\Pi$,
$\phi\in\Gamma(\Sigma,{\cal M}(k^{\otimes-j_\eta}))$ and $m\in I_\eta$,
$$F_{0,\eta,j_\eta}(\phi|R)_{\eta,m}
=\sum_{n=0}^{2j_\eta}{\cal F}^{(j_\eta)}(R){}_m{}^n\partial^n\phi,\eqno(3.15)$$
where, for $w\in\Bbb Z/2$ with $w\geq 0$, ${\cal F}^{(w)}(R)$ is a
$2w+1\times 2w+1$
invertible lower triangular matrix whose entries are differential polynomials
in $R$. From $(3.15)$, $\phi$ being arbitrary, $(3.14)$ follows.
\hfill $\square$

\proclaim 3.
Study of $\Gamma(\Sigma,{\cal O}(k^{\otimes h}\otimes \Ad L))$

\proclaim Theorem 3.5.
Let $\Phi\in\Gamma(\Sigma,{\cal O}(k^{\otimes h}\otimes\Ad L))$. Then, in the
expansion $(3.2)$, one has
$$\eqalignno{\Phi_{\eta,m}=&~0,\quad\eta\in\Pi,~m\in I_\eta,~m>h,&(3.16a)\cr
\Phi_{\eta,h}=&~0,\quad\eta\in\Pi,~h\in I_\eta,~j_\eta>-h.&(3.16b)\cr}$$

\par
{\it Proof}. From $(3.3)$ and $(3.4)$, it follows that, if $\Phi_{\eta,m}=0$
for $m\in I_\eta$, $m>n$ with $n\in I_\eta$, then $\Phi_{\eta,n}\in
\Gamma(\Sigma,{\cal O}(k^{\otimes h-n}))$. On the other hand, $\Gamma(\Sigma,
{\cal O}(k^{\otimes h-n}))=0$ if $h<n$, from the Riemann-Roch theory
\ref{22}. From these properties, beginning with $m=j_\eta$ and proceeding
by induction, one can easily verify $(3.16a)$. If $h\in I_\eta$, then,
from $(3.3)$, $(3.4)$ and $(3.16a)$,
it follows that $\Phi_{\eta,h}\in\Gamma(\Sigma,
{\cal O}(1))$. Thus, $\Phi_{\eta,h}$ is a constant $c_{\eta,h}$ \ref{22}.
Using $(3.3)$, $(3.4)$ and $(3.16a)$ once more, one finds that, when $h>
-j_\eta$, $\Phi_{\eta,h-1\hst1 a}=k_{ab}(\Phi_{\eta,h-1\hst1 b}-C^{+1}_{j_\eta,
h-1}c_{\eta,h}\partial_b\ln k_{ab})$. If $c_{\eta,h}$ were non zero,
$-(C^{+1}_{j_\eta,h-1}c_{\eta,h})^{-1}\Phi_{\eta,h-1}$ would be a
holomorphic connection of the canonical line bundle $k$. These are known not
to exist \ref{22}. Hence, $c_{\eta,h}=0$. \hfill $\square$

\proclaim Definition 3.3.
For any $w\in\Bbb Z/2$ with $w\geq 0$, let
$\{\upsilon^{(w)}_i|i=1,\cdots,d_w\}$
be a basis of $\Gamma(\Sigma,{\cal O}(k^{\otimes w}))$. Further, let $R$ be a
holomorphic projective connection. For any $\eta\in\Pi$ and
$\mu\in I_\eta$ with either $\mu<h$ or $\mu=h=-j_\eta$ and any $i=1,\cdots,~
d_{h-\mu}$, set
$$\Upsilon^{(h)}_{\eta,\mu,i}(R)=F_{h,\eta,\mu}\big(\upsilon^{(h-\mu)}_i|R\big).
\eqno(3.17)$$

\proclaim Theorem 3.6.
For any holomorphic projective connection $R$,
the set $\{\Upsilon^{(h)}_{\eta,\mu,i}(R)|\eta\in\Pi,~\mu\in I_\eta~
{\rm with~either}~\mu<h~{\rm or}~\mu=h=-j_\eta,~i=1,\cdots,d_{h-\mu}\}$
is a basis of $\Gamma(\Sigma,{\cal O}(k^{\otimes h}\otimes\Ad L))$.

\par
{\it Proof}. Since $\upsilon^{(h-\mu)}_i\in\Gamma(\Sigma,{\cal O}
(k^{\otimes h-\mu}))$, $\Upsilon^{(h)}_{\eta,\mu,i}(R)\in\Gamma(\Sigma,
{\cal O}(k^{\otimes h}\otimes\Ad L))$ (cf. th. 3.3). For $\nu\in
\Bbb N\cup\{0\}$, let $\Phi_\nu\in\Gamma(\Sigma,{\cal O}(k^{\otimes h}\otimes
\Ad L))$. Let $\Pi_\nu$ be the subset of $\Pi$ such that, for $\eta\in\Pi_\nu$,
$\Phi_{\nu\hst1\eta,m}\not=0$ for some $m\in I_\eta$. For any $\eta\in
\Pi_\nu$, let $m_{\nu,\eta}$ be the largest value of $m\in I_\eta$ such that
$\Phi_{\nu\hst1\eta,m}\not=0$. By $(3.16)$, either $m_{\nu,\eta}
<h$ or $m_{\nu,\eta}=h=-j_\eta$. From $(3.3)$--$(3.4)$, it follows that
$\phi_{\nu,\eta}\equiv\Phi_{\nu\hst1 \eta,m_{\nu,\eta}}\in\Gamma(\Sigma,{\cal
O}
(k^{\otimes h-m_{\nu,\eta}}))$. Applying th. 3.3, one can construct an
element $F_{h,\eta,m_{\nu,\eta}}(\phi_{\nu,\eta}|R)\in\Gamma(\Sigma,{\cal O}
(k^{\otimes h}\otimes\Ad L))$ satisfying $(3.5)$. Set $\Phi_{\nu+1}=
\Phi_\nu-\sum_{\eta\in\Pi_\nu}F_{h,\eta,m_{\nu,\eta}}(\phi_{\nu,\eta}|R)$.
Clearly, $\Phi_{\nu+1}\in\Gamma(\Sigma,{\cal O}(k^{\otimes h}\otimes
\Ad L))$. It is easily checked that $\Pi_{\nu+1}\subseteq\Pi_\nu$ and that, for
$\eta\in\Pi_{\nu+1}$, $m_{\nu+1,\eta}<m_{\nu,\eta}$. Using the procedure
outlined above, given any element $\Phi\in\Gamma(\Sigma,{\cal O}(k^{\otimes h}
\otimes\Ad L))$, one can construct a finite sequence $\Phi_0,\Phi_1,\cdots,
\Phi_{N+1}$ of elements of $\Gamma(\Sigma,{\cal O}(k^{\otimes h}\otimes\Ad L))$
and a finite sequence $\Pi_0,~\Pi_1,\cdots,~\Pi_{N+1}$ of subsets of $\Pi$
such that $\Phi_0=\Phi$, $\Phi_{N+1}=0$ and $\emptyset=\Pi_{N+1}\subseteq
\Pi_N\subseteq\cdots\subseteq\Pi_0$. In this way, one reaches the
representation
$$\Phi=\sum_{\nu=0}^N\sum_{\eta\in\Pi_\nu}F_{h,\eta,m_{\nu,\eta}}
(\phi_{\nu,\eta}|R).\eqno(3.18)$$
{}From here, it is obvious that the $\Upsilon^{(h)}_{\eta,\mu,i}(R)$ span
$\Gamma
(\Sigma,{\cal O}(k^{\otimes h}\otimes\Ad L))$, since $\phi_{\nu,\eta}$ is
expressible as a linear combination of the $\upsilon^{(h-m_{\nu,\eta})}_i$.
Suppose that $\sum_{\eta,\mu,i}c_{\eta,\mu,i}\Upsilon^{(h)}_{\eta,\mu,i}(R)=0$,
where $c_{\eta,\mu,i}\in\Bbb C$. Then, one also has for each $\eta\in\Pi$,
$\sum_{\mu,i}c_{\eta,\mu,i}\Upsilon^{(h)}_{\eta,\mu,i}(R)_{\eta,m}=0$ for
$m\in I_\eta$. Let $\mu_{0,\eta}$ be the largest value of $\mu$ in the
summation range. Now, by $(3.5a)$ and $(3.17)$, $\Upsilon^{(h)}_{\eta,\mu_{0,
\eta},i}(R)_{\eta,\mu_{0,\eta}}=\upsilon^{(h-\mu_{0,\eta})}_i$. Since the
$\upsilon^{(h)}_i$ are linearly independent $c_{\eta,\mu_{0\eta},i}=0$ for all
$i$. Let $\mu_{1,\eta}$ the next to largest value of $\mu$ in the summation
range. Proceeding as above, one shows that $c_{\eta,\mu_{1,\eta},i}=0$ for all
$i$ and so on. \hfill $\square$

\proclaim Theorem 3.7.
One has
$$\eqalignno{
&\dim\Gamma(\Sigma,{\cal O}(k^{\otimes h}\otimes\Ad L))=
\sum_{\eta\in\Pi,j_\eta\in\{-h,h-1\}}1+\sum_{\eta\in\Pi,j_\eta\in\Bbb Z+h,
j_\eta>\max\{-h,h-1\}}1
&\cr&
+\bigg[\dim\Gamma(\Sigma,{\cal O}(k^{\otimes{1\over 2}}))-{1\over 4}(\ell-1)
\bigg]\cdot
\sum_{\eta\in\Pi,j_\eta\in\Bbb Z+h+{1\over 2},j_\eta>\max\{-h,h-1\}}1
&\cr&
+\bigg[\sum_{\eta\in\Pi,j_\eta>\max\{-h,h-1\}}(j_\eta+h)^2+
\sum_{\eta\in\Pi,-h<j_\eta\leq h-1}(2h-1)(2j_\eta+1)\bigg](\ell-1).~~~
&(3.19)\cr}$$

\par
{\it Proof}. From th. $B.6$, it follows readily from here that
$$\eqalignno{\dim\Gamma(\Sigma &,{\cal O}(k^{\otimes h}\otimes\Ad L))
\cr&
=\sum_{\eta\in\Pi,j_\eta=-h}\dim\Gamma(\Sigma,{\cal O}(1))+
\sum_{\eta\in\Pi,j_\eta>-h,\mu\in I_\eta,\mu<h}\dim\Gamma(\Sigma,
{\cal O}(k^{\otimes h-\mu})).\hskip 1.3truecm &(3.20)}$$
The right hand side of $(3.20)$ can be computed using that $\dim\Gamma(\Sigma,
{\cal O}(k^{\otimes h}))=1,~\ell$ and $(2h-1)(\ell-1)$ respectively for $h=0$,
$h=1$ and $h\geq{3\over 2}$ \ref{22}. The calculation is tedious but
straightforward. \hfill $\square$

\proclaim 4.
Instability of the $G$--bundle $L$

Recall that a holomorphic $G$--bundle $P$ is unstable if $\dim\Gamma(\Sigma,
{\cal O}(\Ad P))>0$. This implies in particular the existence of non trivial
holomorphic gauge transformations of $P$, {\it i. e.} elements of the group
$\Gamma(\Sigma,{\cal O}^*(\Ad P))$ of holomorphic $G$--valued sections of
$\Ad P$.

\proclaim Theorem 3.8.
The $G$--bundle $L$ is unstable.

\par
{\it Proof}. Indeed, from $(3.19)$ with $h=0$, it follows that $\dim\Gamma(
\Sigma,{\cal O}(\Ad L))>0$. \hfill $\square$

\proclaim 5.
Flatness and flat structures of $L$

A holomorphic $G$--bundle $P$ is said flat if it admits a flat form.
Recall that a flat form of a holomorphic $G$--bundle $P$ is a $G$--valued
constant $1$--cocycle $T$ such that $T_{ab}=V_aP_{ab}V_b{}^{-1}$
for some holomorphic $G$--valued $0$--cochain $V$. It can be shown that
$P$ is flat if and only if there is a holomorphic connection of $P$,
{\it i. e.} a holomorphic $\goth g$--valued $0$--cochain $C$ such that
$C_a=k_{ab}(\Ad P_{ab}C_b+\partial_bP_{ab}P_{ab}{}^{-1})$ \ref{23}.
Further, the flat forms of $P$ are in
one-to-one correspondence with the holomorphic gauge equivalence classes of
holomorphic connections of $P$, where the action of a holomorphic gauge
transformation $\gamma\in\Gamma(\Sigma,{\cal O}^*(P))$ on a holomorphic
connection $C$ is given by $\gamma C=\Ad\gamma C+\partial\gamma\gamma^{-1}$
\ref{23}.

\proclaim Theorem 3.9.
The $G$--bundle $L$ is flat.

\par
{\it Proof}. This follows from th. 3.10 below. \hfill $\square$

\proclaim Definition 3.4.
For any holomorphic projective connection $R$, let $A(R)$ be the
$\goth g$--valued $0$--cochain defined by
$$A(R)_a=(1/2)t_{+1}-R_at_{-1},\eqno(3.21)$$

\proclaim Theorem 3.10.
For every holomorphic projective connection $R$, the $\goth g$--valued
$0$--cochain $A(R)$ is a holomorphic connection of $L$.

\par
{\it Proof}. Indeed, using $(2.1)$ and the relation $k_{ab}{}^{-1}
\partial_a{}^2k_{ab}{}^{-1}-{1\over 2}(\partial_ak_{ab}{}^{-1})^2=\{z_a,z_b\}$,
it is straightforward to verify that
$$A(R)_a=k_{ab}(\Ad L_{ab}A(R)_b+\partial_bL_{ab}L_{ab}{}^{-1}),\eqno(3.22)$$
showing the statement. \hfill $\square$

One of the outstanding problems to be tackled is the description of the flat
forms of $L$. I do not have a complete solution of this problem.
The answer is expected to depend on the topology of the group $G$
which the method used here, essentially based on Lie algebra
theory, cannot probe. In spite of this, a number of results can be shown.

\proclaim Definition 3.5.
An element $\Phi\in\Gamma(\Sigma,{\cal O}(\Ad L))$ is said negative graded
if $\Phi$ is valued in $\goth g_{<0}$. A holomorphic gauge
transformation $\gamma\in\Gamma(\Sigma,{\cal O}^*(\Ad L))$ is said negative
graded if it is expressible as $\exp\Theta$ for some negative graded element
$\Theta\in\Gamma(\Sigma,{\cal O}(\Ad L))$.

\proclaim Definition 3.6.
A holomorphic connection $C$ of $L$ is said reduced if, for some holomorphic
projective connection $R$, $C-A(R)$ is valued in
$\ker\ad t_{-1}$, where $A(R)$ is the connection $(3.21)$.

{}From $(3.1)$, one can readily check that these notions are coordinate
independent.

\proclaim Theorem 3.11.
A holomorphic connection $C$ of $L$ is reduced if and only if, for some
holomorphic projective connection $R$, $C$ is of the form
$$C=A(R)+\sum_{\eta\in\Pi}\omega_\eta t_{\eta,-j_\eta},
\quad\omega_\eta\in\Gamma(\Sigma,{\cal O}(k^{\otimes j_\eta+1})).\eqno(3.23)$$
In that case, $C$ admits the above representation for every holomorphic
projective connection $R$.
Hence, the set of reduced holomorphic connections of $L$ can be identified with
the affine space $A(R)+\bigoplus_{\eta\in\Pi}\Gamma(\Sigma,{\cal O}
(k^{\otimes j_\eta+1}))$, the isomorphism depending on the choice of $R$.

\par
{\it Proof}.
$\ker\ad t_{-1}$ is spanned by the generators $t_{\eta,-j_\eta}$
with $\eta\in\Pi$. Thus, $C$ is reduced if and only if it is of the form
$(3.23)$, for some holomorphic projective connection $R$. Now, $C-A(R)\in
\Gamma(\Sigma,{\cal O}(k\otimes\Ad L))$. Using $(3.3)$ and $(3.4)$, one checks
that $\omega_\eta\in\Gamma(\Sigma,{\cal O}(k^{\otimes j_\eta+1}))$.
Finally, from $(3.21)$, it appears that one can change $R$ arbitrarily by
redefining $\omega_o$, where $o\in\Pi$ is defined in sect. 2.
\hfill $\square$

\proclaim Theorem 3.12.
For every holomorphic connection $C$ of $L$, there is a unique negative
graded holomorphic gauge transformation
$\gamma_C\in\Gamma(\Sigma,{\cal O}^*(\Ad L))$ such that the gauge
transformed holomorphic connection $\hat C=\gamma_CC$ is reduced.

\par
{\it Proof}. Pick a holomorphic projective connection $R$.
For any holomorphic connection $C$ of $L$, set
$$\Omega(C|R)=C-A(R).\eqno(3.24)$$
$\Omega(C|R)\in\Gamma(\Sigma,{\cal O}(k\otimes\Ad L))$. If $\gamma\in\Gamma
(\Sigma,{\cal O}^*(\Ad L))$ is a holomorphic gauge transformation, one has
$$\Omega(\gamma C|R)=\Ad\gamma\Omega(C|R)+\partial_{A(R)}\gamma\gamma^{-1},
\eqno(3.25)$$
where $\partial_{A(R)}=\partial-\ad A(R)$ is the covariant derivative
associated
to the connection $A(R)$ acting on $\Gamma(\Sigma,{\cal O}(\Ad L))$.
For $\nu\in\Bbb N\cup\{0\}$, let $\Omega_\nu\in\Gamma(\Sigma,{\cal O}(k\otimes
\Ad L))$ be of the form
$$\Omega_\nu=\sum_{\eta\in\Pi,j_\eta\leq{\nu\over 2}-1}\omega_\eta
t_{\eta,-j_\eta}+{\rm t.o.d.}<-{\nu\over 2}+1,\eqno(3.26)$$
where $\omega_\eta\in\Gamma(\Sigma,{\cal O}(k^{\otimes j_\eta+1}))$ and
the abbreviation
${\rm t.o.d.}<\mu$ denotes terms of $t_0$--degree less than $\mu$. From
$(3.3)$ and $(3.4)$, it follows easily that, for $\eta\in\Pi$ with
$j_\eta\in\Bbb Z+{\nu-1\over 2}$ and $j_\eta\geq{\nu-1\over 2}$,
$\Omega_{\nu\hst1\eta,-{\nu-1\over 2}}$
belongs to $\Gamma(\Sigma,{\cal O}(k^{\otimes{\nu+1\over 2}}))$.
Applying th. 3.3, one can construct
the following negative graded element of $\Gamma(\Sigma,{\cal O}^*(\Ad L))$
$$\gamma_\nu=\exp\Bigg(\sum_{\eta\in\Pi,j_\eta\in\Bbb Z+{\nu-1\over 2},j_\eta
>{\nu-1\over 2}}F_{0,\eta,-{\nu+1\over 2}}(\phi_{\nu,\eta}|R_\nu)\Bigg),\quad
\phi_{\nu,\eta}={2\over C^{+1}_{j_\eta,-{\nu+1\over 2}}}\Omega_{\nu\hst1\eta,
-{\nu-1\over 2}},\eqno(3.27)$$
where $R_\nu$ is any chosen holomorphic projective connection. Define
$$\Omega_{\nu+1}=\Ad\gamma_\nu\Omega_\nu+\partial_{A(R)}\gamma_\nu
\gamma_\nu{}^{-1}.\eqno(3.28)$$
Using the variational formula $\delta e^Xe^{-X}={e^{\ad X}-1\over\ad X}
\delta X$, $(2.2a)$ and $(3.21)$ and $(3.5a)$, one finds
$$\eqalignno{
&\Omega_{\nu+1}=-{1\over 2}\Bigg[t_{+1}, \sum_{\eta\in\Pi,j_\eta\in\Bbb Z
+{\nu-1\over 2},j_\eta>{\nu-1\over 2}}F_{0,\eta,-{\nu+1\over 2}}(\phi_{\nu,
\eta}|R_\nu)_{\eta,-{\nu+1\over 2}}t_{\eta,-{\nu+1\over 2}}\Bigg]
&\cr&
+\sum_{\eta\in\Pi,j_\eta\leq{\nu\over 2}-1}\omega_\eta t_{\eta,-j_\eta}
+\sum_{\eta\in\Pi,j_\eta\in\Bbb Z+{\nu-1\over 2},j_\eta\geq{\nu-1\over 2}}
\Omega_{\nu\hst1\eta,-{\nu-1\over 2}}t_{\eta,-{\nu-1\over 2}}
+{\rm t.o.d.}<-{\nu+1\over 2}+1&\cr
&\hphantom{\Omega_{\nu+1}}
=\sum_{\eta\in\Pi,j_\eta\leq{\nu+1\over 2}-1}\omega_\eta t_{\eta,-j_\eta}
+{\rm t.o.d.}<-{\nu+1\over 2}+1,&\cr
&\hphantom{\Omega_{\nu+1}}\omega_\eta=\Omega_{\nu\hst1\eta,-{\nu-1\over 2}},
\quad\eta\in\Pi,~j_\eta={\nu-1\over 2}.&(3.29)\cr}$$
Thus, $\Omega_{\nu+1}$ is of the form $(3.26)$ with $\nu$ replaced by
$\nu+1$. From $(3.16)$ with $h=1$, every $\Omega\in\Gamma(\Sigma,{\cal O}
(k\otimes\Ad L))$ is of the form $(3.26)$ with $\nu=0$. So, setting
$\Omega_0=\Omega$, one constructs a sequence
$\gamma_0,~\gamma_1,\cdots,~\gamma_N$ of negative graded
holomorphic gauge transformations
of $\Gamma(\Sigma,{\cal O}^*(\Ad L))$ and a sequence $\Omega_1,~\Omega_2,
\cdots,~\Omega_{N+1}$ of elements of $\Gamma(\Sigma,{\cal O}(k\otimes\Ad L))$,
where $N=2j_*+1$, where $j_*$ is defined in sect. 2. From $(3.26)$,
$\Omega_{N+1}$ is of the form
$$\Omega_{N+1}=\sum_{\eta\in\Pi}\omega_\eta t_{\eta,-j_\eta},\eqno(3.30)$$
where $\omega_\eta\in\Gamma(\Sigma,{\cal O}(k^{\otimes j_\eta+1}))$.
Now, take $\Omega=\Omega(C|R)$ and follow the procedure outlined above.
Set
$$\gamma_C=\gamma_N\gamma_{N-1}\cdot\cdots\cdot\gamma_0.\eqno(3.31)$$
Recall that $\goth g_{<0}$
is a nilpotent Lie algebra and that, for a nilpotent Lie algebra, the
Hausdorff--Campbell formula holds with no restriction. From these facts,
it follows that $\gamma_C$ is a negative graded element of
$\Gamma(\Sigma,{\cal O}^*(\Ad L))$. From $(3.25)$ and $(3.28)$, one has
$$\Omega(\gamma_CC|R)=\Omega_{N+1}.\eqno(3.32)$$
Hence, by $(3.24)$ and $(3.30)$, $\gamma_CC$ is reduced. This shows the
existence of $\gamma_C$. Let $\Omega_1,~\Omega_2\in\Gamma(\Sigma,{\cal O}
(k\otimes\Ad L))$ be of the form
$$\Omega_i=\sum_{\eta\in\Pi}\omega_{i\hst1 \eta} t_{\eta,-j_\eta},\eqno(3.33)$$
with $\omega_{i\hst1\eta}\in\Gamma(\Sigma,{\cal O}(k^{\otimes j_\eta+1}))$ and
let $\gamma$ be a negative graded element of $\Gamma(\Sigma,{\cal O}^*(\Ad L))$
such that
$$\Omega_2=\Ad\gamma\Omega_1+\partial_{A(R)}\gamma\gamma^{-1}.\eqno(3.34)$$
$\gamma$ can be written in the form
$$\gamma=\exp\Theta,\eqno(3.35)$$
where $\Theta$ is a negative graded element of $\Gamma(\Sigma,{\cal O}(\Ad
L))$.
Combining $(3.33)$ and $(3.34)$ and using the variational formula
$\delta e^Xe^{-X}={e^{\ad X}-1\over\ad X}\delta X$, one finds
$$\ad t_{-1}\bigg[{\exp\ad\Theta-1\over\ad\Theta}\Big(\partial_{A(R)}\Theta-
[\Omega_1,\Theta]\Big)\bigg]=0.\eqno(3.36)$$
For any $m\in\Bbb Z/2$, let $\pi_m$ the projector on $\goth g_m$
along $\goth g_{\not=m}$. Since $\Theta$ is negative graded, $\pi_m\Theta=0$
for any $m\in\Bbb Z/2$ with $m\geq 0$. Suppose that $\pi_m\Theta=0$ for
all $m\in\Bbb Z/2$ such that $m>n$ where $n\in\Bbb Z/2$ with $n<0$.
By grading reasons, recalling $(3.21)$, $(3.36)$ yields
$$0=-(1/2)[t_{-1},[t_{+1},\pi_n\Theta]]+{\rm t.o.d.}<n.\eqno(3.37)$$
Using that $\ad t_{+1}\goth g\cap\ker\ad t_{-1}=\{0\}$ and that $\goth g_{<0}
\cap\ker\ad t_{+1}=\{0\}$, one concludes that $\pi_n\Theta=0$.
Proceeding by induction, one shows that $\pi_m\Theta=0$ for every $m$.
Thus, $\Theta=0$ and $\gamma=1$. Now, let $\gamma_1$ and $\gamma_2$ be two
negative graded elements of $\Gamma(\Sigma,{\cal O}^*(\Ad L))$ such that
$\gamma_1C$ and $\gamma_2C$ are both reduced. Setting $\Omega_i=
\Omega(\gamma_iC|R)$ and $\gamma=\gamma_2\gamma_1{}^{-1}$ above, $(3.33)$
and $(3.34)$ hold. So, $\gamma_1=\gamma_2$. This shows the uniqueness of
$\gamma_C$. \hfill $\square$

\proclaim Theorem 3.13.
Let $\gamma\in\Gamma(\Sigma,{\cal O}^*(\Ad L))$ be of the form
$\gamma=\exp\Theta$ for some $\Theta\in\Gamma(\Sigma,{\cal O}(\Ad L))$.
Then, $\gamma$ maps the space of reduced holomorphic connections
of $L$ into itself if and only if $\Theta=c$ for some constant
element $c\in\goth c_{\goth s}$.

\par
{\it Proof}. To begin with, one notes that, for any $c\in\goth c_{\goth s}$,
the $\goth g$--valued $0$--cochain $\Theta$ defined by $\Theta_a=c$ belongs
to $\Gamma(\Sigma,{\cal O}(\Ad L))$, as follows easily from $(3.1)$.
Conversely, from $(3.3)$--$(3.4)$ and the fact that the only holomorphic
functions on $\Sigma$ are the constants \ref{22}, one easily shows that,
if $\Theta\in\Gamma(\Sigma,{\cal O}(\Ad L))$
is valued in $\goth c_{\goth s}$, then $\Theta=c$ for some constant element
$c\in\goth c_{\goth s}$. For
$c\in\goth c_{\goth s}$, $\ad c\ker\ad t_{-1}\subseteq\ker\ad t_{-1}$.
So, if $\Theta=c$ for some constant element $c\in\goth c_{\goth s}$,
$\gamma$ maps the space of reduced holomorphic connections of $L$
into itself. This shows sufficiency.
Let $\Omega_1,~\Omega_2\in\Gamma(\Sigma,{\cal O}
(k\otimes\Ad L))$ be of the form $(3.33)$ and suppose that $(3.34)$
holds for some holomorphic projective connection $R$.
Then, $(3.36)$ holds as well. Let $\pi_m$ be defined as below eq.
$(3.36)$. By $(3.16)$, $\pi_m\Theta=\delta_{m,0}c$
for $m\in\Bbb Z/2$ with $m\geq 0$, for some constant
element $c\in\goth c_{\goth s}$. If $\pi_m\Theta=\delta_{m,0}c$
for $m\in\Bbb Z/2$ such that $m>n$ where $n\in\Bbb Z/2$
with $n<0$, then, from $(3.21)$ and $(3.36)$, one gets
$$\eqalignno{0=&~\bigg[t_{-1},{\exp\ad c-1\over\ad c}\big(
\partial c-[\Omega_1,c]-(1/2)[t_{+1},\pi_n\Theta]
\big)+{\rm t.o.d.}<n+1\bigg]&\cr
=&~-(1/2)\bigg[t_{-1},{\exp\ad c-1\over\ad c}
[t_{+1},\pi_n\Theta]\bigg]+{\rm t.o.d.}<n.&(3.38)\cr}$$
Here, I have used that $c$ is constant and that $(\exp\ad c-1)\Omega_1$
is valued in $\ker\ad t_{-1}$. The latter property follows from
the fact that $\Omega_1$ is valued in $\ker\ad t_{-1}$ and the already
mentioned invariance of $\ker\ad t_{-1}$ under $\ad c$.
Reasoning as done below eq. $(3.37)$, one can show that this relation
entails that $\pi_n\Theta=0$. Proceeding by induction, one concludes that
$\pi_m\Theta=\delta_{m,0}c$ for any $m\in\Bbb Z/2$. Thus, $\Theta=c$.
If $C_1$ and $C_2$ are two reduced holomorphic connection of $L$ such that
$\gamma C_1=C_2$, then $\Omega_i=\Omega(C_i|R)$ fulfill the above assumptions.
So $\Theta=c$. This shows necessity. \hfill $\square$
\par\vskip.6cm
\item{4.} {\bf The Drinfeld--Sokolov--Krichever--Novikov spaces
and their properties}
\vskip.4cm
\par
In the first part of this section, I shall review briefly the main properties
of the Krichever--Novikov (KN) spaces, which play an important role in the
geometrical framework expounded below. In the second part, I shall introduce
the Drinfeld--Sokolov--Krichever--Novikov (DSKN) spaces, describe their their
standard bases and study their symmetries.

\proclaim Remark.
Throughout this section, the following assumptions are made.
$\Sigma$ is a compact connected Riemann surface without boundary of genus
$\ell\geq 2$. $k^{\otimes{1\over 2}}$ is a fixed
theta characteristic. $h$ is a fixed element of $\Bbb Z/2$. $G$ is a simple
complex Lie group. $S$ is an $SL(2,\Bbb C)$ subgroup of $G$.

\proclaim 1.
The standard KN theory

The basic ingredients of KN theory are the following:

\par\noindent
$i$) a finite subset $\Delta$ of $\Sigma$ such that $|\Delta|\geq 2$
divided into two disjoint subsets $\Delta_{\rm in}$ and
$\Delta_{\rm out}$ such that $|\Delta_{\rm in}|\geq 1$ and
$|\Delta_{\rm out}|\geq 1$;

\par\noindent
$ii$) an element $\rho$ of $\Gamma(\Sigma,{\cal M}(k))$
holomorphic on $\Sigma\setminus\Delta$ with a simple pole of positive
(negative) residue at each point of $\Delta_{\rm in}$ ($\Delta_{\rm out}$)
and imaginary periods.

\par\noindent
To avoid complication with the Riemann--Roch theory, the points of $\Delta$
will assumed to be in general position \ref{7,24}.

Chosen a base point $P_0\in\Sigma\setminus\Delta$, set
$$t(P)=\real\int_{P_0}^P\rho.\eqno(4.1)$$
$t$ is a singlevalued harmonic function on $\Sigma\setminus\Delta$ with the
property that $t(P)\rightarrow-\infty$ ($+\infty$) when $P$ approaches a
point of $\Delta_{\rm in}$ ($\Delta_{\rm out}$) \ref{7,24}. So, $t$ defines
a notion of euclidean time on $\Sigma$. For any $\tau\in\Bbb R$, the subspace
of $\Sigma$ of time $\tau$ is
$$C_\tau=\{P|P\in\Sigma\setminus\Delta,t(P)=\tau\}.\eqno(4.2)$$
$C_\tau$ is a disjoint union of simple loops in
$\Sigma\setminus\Delta$ for all but finitely many critical values of $\tau$,
whose number is bounded by $2\ell-2+|\Delta|$. The critical values correspond
to processes of topological reconstruction wherewith either one loop splits
into two ore more or two or more loops merge into one. The points of $\Sigma$
where the reconstruction occurs are precisely the zeros of $\rho$ and the
number of loops involved equals the order of the zero plus 2.
For any two $\tau_1,\tau_2\in\Bbb R$,
$C_{\tau_1}$ is homologous to $C_{\tau_2}$ in $\Sigma\setminus\Delta$.
Hence, for any $\omega\in\Gamma(\Sigma,{\cal M}(k))$ with poles contained
in $\Delta$, $\oint_{C_\tau}\omega$ is $\tau$--independent.

The KN space $\sans K\sans N_h(\Delta)$ of weight $h$ is
the set of the elements of $\Gamma(\Sigma,{\cal M}(k^{\otimes h}))$
whose poles are contained in $\Delta$
\footnote{${}^1$}{In a different definition, also essential
singularities at the points of $\Delta$ are allowed.}.
$\sans K\sans N_h(\Delta)$ is an infinite dimensional complex vector space.

There exists a bilinear pairing of the spaces
$\sans K\sans N_h(\Delta)$ and
$\sans K\sans N_{1-h}(\Delta)$ defined by
$$\langle\psi,\phi\rangle=\oint_{C_\tau}{dz\over 2\pi i}\phi\psi\eqno(4.3)$$
for any $\phi\in\sans K\sans N_h(\Delta)$ and
$\psi\in\sans K\sans N_{1-h}(\Delta)$.
Note that the integration is well-defined and independent from $\tau\in\Bbb R$.

The space $\sans K\sans N_h(\Delta)$
possesses a standard basis, the generalized KN basis.
To describe this, set $r=|\Delta_{\rm in}|$, $s=|\Delta_{\rm out}|$
and $p_h=p_{1-h}=h-[h]~{\rm mod}~\Bbb Z$. The basis is of the form
$\{\upsilon^{(h)}_{i,N}|i=1,\cdots,r,~N\in\Bbb Z+p_h\}$. The basis
elements $\upsilon^{(h)}_{i,N}$ are characterized up to normalization
by their zero order at the points $\Delta$. Let $\Delta_{\rm in}=
\{P_j|1\leq j\leq r\}$ and $\Delta_{\rm out}=\{P_j|r+1\leq j\leq r+s\}$.
Then,
$$\ord\upsilon^{(h)}_{i,N}(P_j)=a_j(N+1-h)-\delta_{j,i}
+(2h-1)(\ell-1)\delta_{j,r+s}+b_{j,N},\eqno(4.4a)$$
where
$$a_j=\cases{1,& $1\leq j\leq r$,\cr -1,& $r+1\leq j\leq r+\min(r,s)-1$,\cr
-(|r-s|+1)^{\sgn(r-s)},& $r+\min(r,s)\leq j\leq r+s$,\cr}\eqno(4.4b)$$
and the $b_{j,N}$ are rational numbers such that
$$\sum_{j=1}^{r+s}b_{j,N}=0,\quad |b_{j,N}|<1,\quad
b_{j,N}=0,\quad 1\leq j\leq r+\min(r,s)-1\eqno(4.4c)$$
depending on $j$ and $N$. These statements about $b_{j,N}$ must be amended
for finitely may values of $N$ when $h$ takes the exceptional values $0$,
$1\over 2$ for an odd theta characteristic and $1$. See ref. \ref{24} for
a detailed treatment of this matter and refs. \ref{25--28} for related
approaches to the subject.

The relative normalization of the elements
of the KN bases of $\sans K\sans N_h(\Delta)$ and
$\sans K\sans N_{1-h}(\Delta)$ can be chosen so that
$$\langle\upsilon^{(1-h)}_{i,M},\upsilon^{(h)}_{j,N}\rangle
=\delta_{i,j}\delta_{M,-N},\quad i,j=1,\cdots,r,~M,N\in\Bbb Z+p_h.\eqno(4.5)$$
The Laurent theorem generalizes and one gets the expansion
$$\phi=\sum_{i=1}^r\sum_{N\in\Bbb Z+p_h}\phi_{i,N}\upsilon^{(h)}_{i,-N},
\quad\phi_{i,N}=\langle\upsilon^{(1-h)}_{i,N},\phi\rangle,\eqno(4.6)$$
the series containing only a finite number of non vanishing terms
\ref{24}. $(4.5)$ and $(4.6)$ imply further that the pairing
$(4.3)$ is non singular and that the spaces
$\sans K\sans N_h(\Delta)$ and
$\sans K\sans N_{1-h}(\Delta)$ are
reciprocally dual.

The basic symmetry group of the KN theory is the conformal group
$\Conf_0(\Delta)$, {\it i. e.} the group of holomorphic
diffeomorphisms $f$ of $\Sigma\setminus\Delta$ onto itself with holomorphic
inverse having finite order singularities at the points of $\Delta$
and homotopic to $\id_\Sigma$. Its Lie algebra is
$\Lie\Conf_0(\Delta)=\sans K\sans N_{-1}(\Delta)$.
The Lie brackets are given by
$$[u,v]=u\partial v-v\partial u\eqno(4.7)$$
for any two $u,v\in\Lie\Conf_0(\Delta)$.

$\Conf_0(\Delta)$ acts on the KN spaces
$\sans K\sans N_h(\Delta)$.
For any $f\in\Conf_0(\Delta)$ and $\phi\in
\sans K\sans N_h(\Delta)$,
the action is defined by
$$f^*\phi_a=k^{\otimes h}(f)_{ab}\phi_b\circ f.\eqno(4.8)$$
Here,
$$k^{\otimes{1\over 2}}(f)_{ab}=(\partial_af_b)^{1\over 2},\eqno(4.9)$$
where the branch of the square root is chosen so that
$$k^{\otimes{1\over 2}}(\id_\Sigma)_{ab}=k^{\otimes{1\over 2}}{}_{ab},
\eqno(4.10a)$$
$$k^{\otimes{1\over 2}}(g\circ f)_{ac}=
k^{\otimes{1\over 2}}(f)_{ab}k^{\otimes{1\over 2}}(g)_{bc}\circ
f,\eqno(4.10b)$$
for $f,g\in\Conf_0(\Delta)$.
$k^{\otimes h}(f)=k^{\otimes{1\over 2}}(f)^{\otimes 2h}$, by definition.
At the infinitesimal level, $(4.8)$ reduces into
$$\theta_u\phi=u\partial\phi+h(\partial u)\phi\eqno(4.11)$$
for any $u\in\Lie\Conf_0(\Delta)$.
The KN pairing $(4.3)$ is invariant under $\Conf_0(\Delta)$:
$$\langle f^*\psi,f^*\phi\rangle=\langle\psi,\phi\rangle,\eqno(4.12)$$
$$\langle\theta_u\psi,\phi\rangle+\langle\psi,\theta_u\phi\rangle=0
\eqno(4.13)$$
for any $f\in\Conf_0(\Delta)$,
$u\in\Lie\Conf_0(\Delta)$ and any
$\phi\in\sans K\sans N_h(\Delta)$ and
$\psi\in\sans K\sans N_{1-h}(\Delta)$.

\proclaim 2.
The DSKN spaces

\proclaim Definition 4.1.
The DSKN space $\sans D\sans S_h(\Delta)$ of weight $h$ is the set of the
elements of $\Gamma(\Sigma,{\cal M}(k^{\otimes h}\otimes\Ad L))$ whose
poles are contained in $\Delta$.

$\sans D\sans S_h(\Delta)$ is an infinite dimensional complex vector space.

\proclaim Definition 4.2.
For any $\Phi\in\sans D\sans S_h(\Delta)$ and $\Psi\in\sans D\sans S_{1-h}
(\Delta)$, set
$$\langle\Psi,\Phi\rangle=\oint_{C_\tau}{dz\over 2\pi i}(\Phi,\Psi).
\eqno(4.14)$$

Note that the integration is well-defined and independent from $\tau\in\Bbb R$.
$(4.14)$ defines a bilinear pairing of the spaces $\sans D\sans S_h(\Delta)$
and $\sans D\sans S_{1-h}(\Delta)$, called the DSKN pairing. It the appropriate
generalization of the customary KN pairing in the present context.

\proclaim 3.
The DSKN bases

The space $\sans D\sans S_h(\Delta)$ admits standard bases. To construct them,
one needs the following result.

Recall that a meromorphic connection $\varpi$ of $k$ on $\Sigma$
is a meromorphic $0$--cochain $\{\varpi_a\}$ on $\Sigma$ such
that $\varpi_a=k_{ab}(\varpi_b+\partial_b\ln k_{ab})$.

\proclaim Definition 4.3.
Let $\varpi$ be a meromorphic connection of $k$ on $\Sigma$ whose poles
are contained in $\Delta$. Let further $\eta\in\Pi$, $\mu\in I_\eta$ and
$\phi\in\sans K\sans N_{h-\mu}(\Delta)$. One sets
$$Q_{h,\eta,\mu}(\phi|\varpi)=\exp(-\varpi\ad t_{-1})\phi t_{\eta,\mu}.
\eqno(4.15)$$

\proclaim Theorem 4.1.
For any meromorphic connection $\varpi$ of $k$ on $\Sigma$ whose poles
are contained in $\Delta$ and any $\eta\in\Pi$, $\mu\in I_\eta$ and
$\phi\in\sans K\sans N_{h-\mu}(\Delta)$,
$Q_{h,\eta,\mu}(\phi|\varpi)\in\sans D\sans S_h(\Delta)$.
Further, $Q_{h,\eta,\mu}(\phi|\varpi)$ depends linearly on $\phi$.

\par
{\it Proof}. Indeed, using $(2.1)$, one has
$$\exp(-\varpi_a\ad t_{-1})t_{\eta,\mu}=
k^{\otimes\mu}{}_{ab}\Ad L_{ab}\Big(\exp(-\varpi_b\ad t_{-1})t_{\eta,\mu}
\Big).\eqno(4.16)$$
{}From here, the statement is obvious. \hfill $\square$

\proclaim Definition 4.4.
Let $\varpi$ be a meromorphic connection of $k$ on $\Sigma$ whose poles
are contained in $\Delta$. For any $\eta\in\Pi$, $\mu\in I_\eta$, $i=1,
\cdots,~r$ and $N\in\Bbb Z+p_h$, set
$$\Upsilon^{(h)}_{\eta,\mu,i,N}(\varpi)=Q_{h,\eta,\mu}
\big(\upsilon^{(h-\mu)}_{i,N}|\varpi\big).\eqno(4.17)$$

\proclaim Theorem 4.2.
Let $\varpi$ be a meromorphic connection of $k$ on $\Sigma$ whose poles
are contained in $\Delta$.
Then, the set $\{\Upsilon^{(h)}_{\eta,\mu,i,N}(\varpi)|\eta\in\Pi,~\mu\in
I_\eta,~i=1,\cdots,~r,~N\in\Bbb Z+p_h\}$ is a basis of
$\sans D\sans S_h(\Delta)$.
The basis elements satisfy the relation
$$\eqalignno{\langle\Upsilon^{(1-h)}_{\eta,\mu,i,M}(\varpi),
&\Upsilon^{(h)}_{\zeta,\nu,j,N}(\varpi)\rangle
=N_\eta\delta_{\eta,\bar\zeta}(-1)^{j_\eta-\mu}\delta_{\mu,-\nu}
\delta_{i,j}\delta_{M,-N},&\cr&
\eta,\zeta\in\Pi,~\mu,\nu\in I_\eta,~i,j=1,\cdots,~r,~M,N\in\Bbb Z+p_h.
&(4.18)\cr}$$
Finally, for any $\Phi\in\sans D\sans S_h(\Delta)$, one has
$$\eqalignno{
\Phi=&~\sum_{\eta\in\Pi,\mu\in I_\eta}\sum_{i=1}^r\sum_{N\in\Bbb Z+p_h}
\Phi_{\eta,\mu,i,N}(\varpi)\Upsilon^{(h)}_{\bar\eta,-\mu,i,-N}(\varpi),&\cr
\Phi_{\eta,\mu,i,N}(\varpi)=&~\langle\Upsilon^{(1-h)}_{\eta,\mu,i,N}(\varpi),
\Phi\rangle/N_\eta(-1)^{j_\eta-\mu},&(4.19)\cr}$$
the series containing only a finite number of non vanishing terms.

\par
{\it Proof}. Let $\Phi\in\sans D\sans S_h(\Delta)$.
Following a procedure totally analogous to that leading to
$(3.18)$, one shows that
$$\Phi=\sum_{\nu=0}^K\sum_{\eta\in\Pi_\nu}Q_{h,\eta,m_{\nu,\eta}}
(\phi_{\nu,\eta}|\varpi),\eqno(4.20)$$
where $K$ is some non negative integer, $\Pi_\nu$ is a subset of $\Pi$
and $\phi_{\nu,\eta}\in\sans K\sans N_{h-m_{\nu,\eta}}(\Delta)$
for each $\nu$ and $\eta$. Each $\phi_{\nu,\eta}$ is given by a
series of the form $(4.6)$. Hence, the $\Upsilon^{(h)}_{\eta,\mu,i,N}(\varpi)$
span $\sans D\sans S_h(\Delta)$. The
linear independence of the $\Upsilon^{(h)}_{\eta,\mu,i,N}(\varpi)$ is
equivalent to that of the fields $\upsilon^{(h)}_{i,N}t_{\eta,\mu}$ which
is obvious. $(4.18)$ is a straightforward consequence of $(4.14)$, $(4.5)$
and $(2.10)$. $(4.19)$ follows from the representation $(4.18)$ and
from $(4.6)$. \hfill $\square$

\proclaim Theorem 4.3.
The pairing $(4.14)$ is non singular. Therefore, the spaces
$\sans D\sans S_h(\Delta)$ and $\sans D\sans S_{1-h}(\Delta)$
are reciprocally dual.

\par
{\it Proof}. This follows directly from $(4.19)$ and $(4.20)$.
\hfill $\square$

\proclaim 4.
The symmetries of the DSKN spaces

There exists a natural extension of the action of $\Conf_0(\Delta)$ to the
DSKN spaces. This leads to the following result.

\proclaim Theorem 4.4.
$\Conf_0(\Delta)$ acts on the space $\sans D\sans S_h(\Delta)$ by setting
$$f^*\Phi_a=k^{\otimes h}(f)_{ab}\Ad L(f)_{ab}\Phi_b\circ f
\eqno(4.21)$$
for arbitrary $f\in\Conf_0(\Delta)$ and $\Phi\in\sans D\sans S_h(\Delta)$,
where,
$$L(f)_{ab}=\exp(-2\ln k^{\otimes{1\over 2}}(f)_{ab}t_0)
\exp(\partial_ak(f)_{ab}{}^{-1}t_{-1}).\eqno(4.22)$$
At the infinitesimal level, $(4.21)$ can be written as
$$\theta_u\Phi=u\partial_{A(R)}\Phi
+h(\partial u)\Phi+[\dot L(u|R),\Phi]\eqno(4.23)$$
for any $u\in\Lie\Conf_0(\Delta)$, where
$R$ is a holomorphic projective connection and
$$\dot L(u|R)=\big[(1/2)t_{+1}-\partial t_0-(\partial^2+R)t_{-1}\big]u,
\eqno(4.24)$$
$\partial_{A(R)}=\partial-\ad A(R)$ being the covariant derivative of the
connection $A(R)$ defined in $(3.21)$.
$\dot L(u|R)\in\sans D\sans S_0(\Delta)$ and $\dot L(u|R)$
satisfies the equation
$$\partial_{A(R)}\dot L(u|R)=-D_1(R)ut_{-1},\quad D_1(R)=\partial^3+2R\partial
+(\partial R),\eqno(4.25)$$
where $D_1(R)$ is a Bol operator $\ref{21}$.

\par
{\it Proof}. From $(4.10a)$, it is easily verified that
$$L(\id_\Sigma)_{ab}=L_{ab}.\eqno(4.26a)$$
{}From $(2.1)$ and $(4.10b)$, one verifies further that
$$L(g\circ f)_{ac}=L(f)_{ab}L(g)_{bc}\circ f\eqno(4.26b)$$
for any $f,g\in\Conf_0(\Delta)$.
Using $(4.10)$ and $(4.26)$ in combination, it is
straightforward to verify that the right hand side of $(4.21)$ belongs
to $\sans D\sans S_h(\Delta)$ the remaining statements are straightforwardly
verified. \hfill $\square$

Note that $\theta_u\Phi$ is independent from $R$. $R$ is introduced only in
order the various contributions appearing in its expression to have nice
covariance properties.

\proclaim Theorem 4.5.
The pairing $(4.14)$ is invariant under $\Conf_0(\Delta)$. In fact, one has
$$\langle f^*\Psi,f^*\Phi\rangle=\langle\Psi,\Phi\rangle,\eqno(4.27)$$
$$\langle\theta_u\Psi,\Phi\rangle+\langle\Psi,\theta_u\Phi\rangle=0
\eqno(4.28)$$
for any $f\in\Conf_0(\Delta)$, $u\in\Lie\Conf_0(\Sigma
\setminus\Delta)$ and any $\Phi\in\sans D\sans S_h(\Delta)$ and
$\Psi\in\sans D\sans S_{1-h}(\Delta)$.

\par
{\it Proof}. The verification is straightforward.\hfill $\square$

Denote by $\Gau_0(\Delta)$ the group of gauge transformations
$\gamma\in\Gamma(\Sigma\setminus\Delta,{\cal M}^*(\Ad L))$ with
finite order singularities at the points of $\Delta$ and homotopic to the
identity. Its Lie algebra $\Lie\Gau_0(\Delta)$ is
$\sans D\sans S_0(\Delta)$ with Lie brackets
$$[\Xi,\Lambda]=[e(\Xi),e(\Lambda)]\eqno(4.29)$$
for any two $\Xi,\Lambda\in\Lie\Gau_0(\Delta)$,
where in the right hand side $e$ is the evaluation map at a given point of
$\Sigma\setminus\Delta$ and the Lie brackets are those of $\goth g$.

$\Gau_0(\Delta)$ acts on the DSKN spaces by means of the adjoint
representation, as summarized by the following theorem.

\proclaim Theorem 4.6.
$\Gau_0(\Delta)$ acts on he space $\sans D\sans S_h(\Delta)$ by setting
$$\gamma\Phi=\Ad\gamma\Phi\eqno(4.30)$$
for arbitrary $\gamma\in\Gau_0(\Delta)$ and $\Phi\in\sans D\sans S_h(\Delta)$.
At the infinitesimal level, $(4.30)$ becomes
$$\delta_\Xi\Phi=[\Xi,\Phi]\eqno(4.31)$$
for any $\Xi\in\Lie\Gau_0$.

\par
{\it Proof}. The verification is straightforward.\hfill $\square$

\proclaim Theorem 4.7.
The pairing $(4.14)$ is invariant under $\Gau_0(\Delta)$. In fact, one has
$$\langle\gamma\Psi,\gamma\Phi\rangle=\langle\Psi,\Phi\rangle,\eqno(4.32)$$
$$\langle\delta_\Xi\Psi,\Phi\rangle+\langle\Psi,\delta_\Xi\Phi\rangle=0
\eqno(4.33)$$
for any $\gamma\in\Gau_0(\Delta)$,
$\Xi\in\Lie\Gau_0(\Delta)$ and any $\Phi\in\sans D
\sans S_h(\Delta)$ and and $\Psi\in\sans D\sans S_{1-h}(\Delta)$.

\par
{\it Proof}. This verification is also straightforward.\hfill $\square$

Note that the total symmetry group is the semidirect product
$\Conf_0(\Delta)\ltimes\Gau_0(\Delta)$
where the product is defined by $(f_1,\gamma_1)\circ(f_2,\gamma_2)=
(f_1\circ f_2,\gamma_1f_1{}^{-1*}\gamma_2)$, for $f_1,f_1\in
\Conf_0(\Delta)$ and $\gamma_1,\gamma_2\in\Gau_0(\Sigma
\setminus\Delta,L)$. The action of the first and second factor
are respectively right and left.
\par\vskip.6cm
\item{5.} {\bf The Poisson manifold} $(\sans W,\{\cdot,\cdot\}_\kappa)$
\vskip.4cm
\par
In this section, I shall introduce a Poisson manifold $(\sans W,
\{\cdot,\cdot\}_\kappa)$ which is closely related to the customary
Kac--Moody phase space, though its geometry is some respect quite different.
In fact, the construction uses the DS bundle $L$ and the DSKN spaces
in an essential way.

\proclaim Remark. In this section,
$\Sigma$ is a compact connected Riemann surface without boundary of genus
$\ell\geq 2$.  $G$ is a simple complex Lie group. $S$ is an $SL(2,\Bbb C)$
subgroup of $G$.

In the application of the KN theory below, $\Delta$ consists of just two points
$P_+$ and $P_-$ in general position. $\Delta_{\rm in}$ and $\Delta_{\rm out}$
contain respectively the point $P_+$ and $P_-$.
$\rho$ is the unique element of $\Gamma(\Sigma,{\cal M}(k))$
holomorphic on $\Sigma\setminus\Delta$ with a simple pole of
residue $+1$ ($-1$) at $P_+$ ($P_-$) and imaginary periods.

To lighten the notation, the dependence of the functional spaces encountered
below on $\Delta$ will be omitted.

The relevant space of the construction is
$$\sans W=\sans D\sans S_1,\eqno(5.1)$$
that is the the DSKN space of weight $1$.
$\sans W$ is an infinite dimensional complex vector space
and, thus, also an infinite dimensional holomorphic manifold.
The relevant function space on $\sans W$ is the space $\goth D(\sans W)$
of differential polynomials on $\sans W$.
$\sans W$ can be endowed with a Poisson structure depending on a parameter
$\kappa\in{\Bbb C}\setminus\{0\}$ and supported on $\goth D(\sans W)$.
The Poisson structure is completely defined by assigning the Poisson
brackets of the linear inhomogeneous functionals on $\sans W$.
The Poisson brackets of general elements of $\goth D(\sans W)$ are obtained
by enforcing the Leibniz rule. This leads to considering the dual space
$\sans W^\vee$ of $\sans W$. Under the non singular DSKN pairing $(4.14)$,
one has the identification
$$\sans W^\vee=\sans D\sans S_0.\eqno(5.2)$$
Therefore, every linear functional on $\sans W$ is of the form
$$\lambda_X(W)=\langle X,W\rangle,\quad W\in\sans W,\eqno(5.3)$$
for some $X\in\sans W^\vee$.
Note that $\sans W^\vee$ has an obvious structure of Lie algebra.

For any $X,~Y\in\sans W^\vee$ and any $a,~b\in{\Bbb C}$,
the Poisson brackets of the inhomogeneous linear functionals
$\lambda_X+a$ and $\lambda_Y+b$  are given by
$$\{\lambda_X+a,\lambda_Y+b\}_\kappa=\lambda_{[X,Y]}+\kappa\chi(X,Y),
\eqno(5.4a)$$
$$\chi(X,Y)=\langle X,\partial_AY\rangle,\eqno(5.4b)$$
where $A$ is the holomorphic connection of $L$ given by $(3.21)$ for
some holomorphic projective connection $R$. $R$ will be fixed
once for all in the following. So, the dependence
on $R$ will be understood below, to simplify the notation.

It is straightforward to verify that the Poisson brackets
$\{\cdot,\cdot\}_\kappa$ are bilinear, antisymmetric and satisfy the Jacobi
identity as they should. In fact, one easily checks that $\chi$ is a
Lie algebra $1$--cocycle of $\sans W^\vee$. $\chi$ depends on the choice of
$R$, but changing the choice alters $\chi$ by a trivial $1$--cocycle.
$\chi$ is singular, since $\chi(X,Y)=0$ identically whenever either $X$ or
$Y$ are constant elements of $\goth c_{\goth s}$.

The above Poisson structure provides the proper geometric definition
of Kac--Moody phase space in the present context. The level and the
Kac--Moody current correspond to $-\kappa$ and $\kappa A+W$, respectively.

Next, one has to consider the symmetries of the Poisson manifold $\sans W$.
These are given by suitable deformations of the conformal and gauge
symmetries introduced in sect. 4.

Consider $\Conf_0$. For any $f\in\Conf_0$,
one has that $\chi(f^*X,f^*Y)=\chi(X,Y)-\kappa\langle [X,Y],A-f^{-1*}A\rangle$,
where for $f\in\Conf_0$, $f^*A=\partial L_fL_f{}^{-1}+\partial f
\Ad L_fA\circ f$ and $X,Y\in\sans W^\vee$. Because of the non invariance of
$\chi$, the action of $\Conf_0$ on $\sans W$, defined by
$(4.21)$, is not Poisson: it does not leave the Poisson brackets invariant.
However, there is a deformation of the action enjoying this property. Set
$$(f^*W)_\kappa=f^*W+\kappa(f^*A-A), \quad W\in\sans W.\eqno(5.5)$$
The deformation induces an action of $\Conf_0$ on
$\goth D(\sans W)$. It is sufficient to consider the action
on the functionals $\lambda_X+a$, $X\in\sans W^\vee$, $a\in{\Bbb C}$,
which is given by
$$(f^*(\lambda_X+a))_\kappa(W)=\lambda_X((f^{-1*}W)_\kappa)+a
=\lambda_{f^*X}(W)+a+\kappa\langle X,f^{-1*}A-A\rangle,
\quad W\in\sans W.\eqno(5.6)$$
{}From $(5.4)$ and $(5.6)$, it follows that the action $(5.5)$ is Poisson.

At the infinitesimal level, $(5.5)$ and $(5.6)$ become
$$(\theta_uW)_\kappa=\kappa\partial_A\dot L(u)+\theta_uW,\eqno(5.7)$$
$$(\theta_u(\lambda_X+a))_\kappa(W)=-\lambda_X((\theta_uW)_\kappa)=
\lambda_{\theta_uX}(W)+\kappa\langle \dot L(u),\partial_A X\rangle,\eqno(5.8)$$
where $u\in\Lie\Conf_0$ and $\dot L(u)$ is given by $(4.24)$.
The action is Hamiltonian. In fact, using $(5.5)$, it is straightforward
to verify that
$$(\theta_u(\lambda_X+a))_\kappa=\{T_u,\lambda_X+a\}_\kappa,\eqno(5.9a)$$
$$T_u(W)=(1/2\kappa)\langle uW,W\rangle+\langle \dot L(u), W\rangle,
\quad W\in\sans W,\eqno(5.9b)$$
the Hamiltonian functions $T_u$ being elements of $\goth D(\sans W)$.
$T_u$ can be written as
$$T_u(W)=\langle u,T(W)\rangle,\quad W\in\sans W\eqno(5.10a)$$
$$T(W)=\big((1/2)t_{+1}+\partial t_0-(\partial^2+R)t_{-1},W\big)
+(1/2\kappa)(W,W).\eqno(5.10b)$$
Using $(3.1)$ and $(2.1)$ and $(2.9)$, its is straightforward albeit
lengthy to check that $T(W)\in\sans K\sans N_2$.
So, the map $W\in\sans W\rightarrow T(W)\in\sans K\sans N_2$ is the
moment map of the Hamiltonian action.

For any $\gamma\in\Gau_0$, one has that $\chi(\gamma X,
\gamma Y)=\chi(X,Y)-\langle [X,Y],\gamma^{-1}\partial_A\gamma\rangle$.
So, the ordinary action of $\Gau_0$ on $\sans W$, defined
by $(4.30)$, is not Poisson. However, in this case too, there exists a
deformation of the action enjoying such property, namely
$$(\gamma W)_\kappa=\gamma W+\kappa\partial_A\gamma\gamma^{-1},
\quad W\in\sans W.\eqno(5.11)$$
The deformation induces an action of $\Gau_0$
on the functionals $\lambda_X+a$, $X\in\sans W^\vee$,
$a\in{\Bbb C}$:
$$(\gamma(\lambda_X+a))_\kappa(W)=\lambda_X((\gamma^{-1}W)_\kappa)+a
=\lambda_{\gamma X}(W)+a+\kappa\langle X,\partial_A\gamma^{-1}\gamma
\rangle,\quad W\in\sans W.\eqno(5.12)$$
By combining $(5.4)$ and $(5.12)$, one verifies that the deformed action thus
defined is Poisson. At the infinitesimal level, $(5.11)$ and $(5.12)$ become
$$(\delta_\Xi W)_\kappa=\delta_\Xi W+\kappa\partial_A\Xi,\eqno(5.13)$$
$$(\delta_\Xi(\lambda_X+a))_\kappa(W)=-\lambda_X((\delta_\Xi W)_\kappa)=
\lambda_{\delta_\Xi X}(W)+\kappa\langle\Xi,\partial_A X\rangle,\eqno(5.14)$$
where $\Xi\in\Lie\Gau_0$ (cf. eq. $(4.31)$).
{}From $(5.4)$ and $(5.14)$, one has
$$(\delta_\Xi(\lambda_X+a))_\kappa=\{J_\Xi,\lambda_X+a\}_\kappa,\eqno(5.15a)$$
$$J_\Xi(W)=\langle\Xi,W\rangle,\quad W\in\sans W.\eqno(5.15b)$$
Note that $J_\Xi\in\goth D(\sans W)$.
{}From here, it appears that the deformed action of $\Gau_0$ on
$\sans W$ is Hamiltonian with respect to the Poisson structure $(5.4)$,
the Hamiltonian functions being the $J_\Xi$.
$J_\Xi$ can trivially be written as
$$J_\Xi(W)=\langle\Xi,J(W)\rangle,\quad W\in\sans W.\eqno(5.16a)$$
$$J(W)=W.\eqno(5.16b)$$
So, the map $W\in\sans W\rightarrow J(W)\in\sans W$ can be identified
with the moment map of the Hamiltonian action.

By a straightforward calculation, one obtains
$$\{T_u,T_v\}_\kappa=T_{[u,v]}+12\kappa(t_0,t_0)\sigma(u,v),
\quad u,v\in\Lie\Conf_0,\eqno(5.17a)$$
where
$$\sigma(u,v)=-{1\over 12}\langle u,D_1v\rangle\eqno(5.17b)$$
is the KN $1$--cocycle and $D_1$ is given in $(4.25)$.
The proof of $(5.17)$ uses $(5.4)$, $(5.9b)$, $(2.9)$
and the following two relations
$$u\partial_A\dot L(v)-v\partial_A\dot L(u)+[\dot L(u),\dot L(v)]
=\dot L([u,v]),\eqno(5.18)$$
$$\chi(\dot L(u),\dot L(v))=-(t_0,t_0)\langle u,D_1v\rangle,\eqno(5.19)$$
which are easily verified using $(2.1)$ and $(4.24)$-$(4.25)$.
$(5.17)$ is a Poisson bracket Virasoro algebra of central charge
$12\kappa(t_0,t_0)$. This is the well-known value of the classical
central charge encountered in the theory of classical $W$--algebras
\ref{1,9--13}. The moment map $T(W)$, eq. $(5.10b)$, is the
energy-momentum tensor. In the usual approach \ref{9--13},
the central charge originates from an improvement term added to the Sugawara
energy-momentum tensor of Kac--Moody theory in order to maintain conformal
invariance upon carrying out the Hamiltonian reduction of the Kac--Moody
phase space. The first and second contributions in expression $(5.9b)$ of
$T_v$ correspond more or less to such terms in the present formulation.
Here, however, the improvement term is yielded {\it ab initio} by
the nature of the DS vector bundle and the action of
the conformal group of $\Sigma\setminus\Delta$. The second derivative term
appearing in expression $(5.10b)$ of $T(W)$ has a counterpart in the usual
approach where it is added {\it ad hoc} after the reduction of the phase
space \ref{9--13}. Here, it is present from the beginning and it is
strictly necessary to ensure the correct transformation properties of $T(W)$
under coordinate changes.

{}From $(5.4)$ and $(5.15b)$, one gets
$$\{J_\Xi,J_\Lambda\}_\kappa=J_{[\Xi,\Lambda]}+\kappa\chi(\Xi,\Lambda),
\quad \Xi,\Lambda\in\Lie\Gau_0,\eqno(5.20)$$
$(5.20)$ is a Poisson bracket Kac--Moody algebra of level $\kappa$. The
moment map $J(W)$, eq. $(5.16b)$, plays here the role of the Kac--Moody
current.

{}From $(5.4)$, $(5.9b)$ and $(5.15b)$, one also has
$$\{T_u,J_\Xi\}_\kappa=J_{\theta_u\Xi}+\kappa\chi(\dot L(u),\Xi),
\quad u\in\Lie\Conf_0,~\Xi\in\Lie\Gau_0,\eqno(5.21)$$
Hence, the current $J(W)$ transforms as a primary field under
Poisson bracketting, except for the component corresponding to the generator
$t_{+1}$ of $\goth g$ (see eqs. $(4.25)$ and $(5.4b)$). This also is familiar
in the theory of classical $W$--algebras \ref{9--13}.

It is interesting to write the Poisson bracket algebra in modes. One uses the
KN $\{\upsilon_{i,N}^{(h)}\}$ and the DSKN bases
$\{\Upsilon_{\eta,\mu,i,N}^{(h)}(\varpi)\}$ introduced in sect. 4, where
$\varpi$ is a meromorphic connection of $k$ holomorphic on
$\Sigma\setminus\Delta$.
In this case, $r$ being $1$, one can suppress the index $i$.
To simplify the notation, the dependence on $\varpi$ will be understood. Set
$$T_P=T_{\upsilon^{(-1)}_P},\eqno(5.22)$$
$$J_{\eta,\mu,M}=J_{\Upsilon_{\eta,\mu,M}^{(0)}},\eqno(5.23)$$
for $P\in\Bbb Z$ and $\eta\in\Pi$, $\mu\in I_\eta$ and $M\in\Bbb Z+p_{j_\eta}$.
Then, $(5.17)$--$(5.21)$ yield the following algebra
$$\{T_P,T_Q\}_\kappa=\sum_{R\in \Bbb Z}c^{(-1)}_P{}^{(-1)}_Q{}^R_{(-1)}T_R
-\kappa(t_0,t_0)\varsigma^{(-1)}_P{}^{(-1)}_Q,\eqno(5.24)$$
$$\eqalignno{
\{J_{\eta,\mu,M},J_{\zeta,\nu,N}\}_\kappa=&\sum_{\xi\in\Pi}
\sum_{L\in\Bbb Z+p_{j_\xi}}F_{\eta,\zeta}{}^\xi(j_\eta,\mu;j_\zeta,\nu|j_\xi,
\mu+\nu)f^{(-\mu)}_M{}^{(-\nu)}_N{}^L_{(-\mu-\nu)}J_{\xi,\mu+\nu,L}&\cr
&+\kappa N_\eta\delta_{\eta,\bar\zeta}(-1)^{j_\eta-\mu}\Big[
-(1/2)C^{+1}_{j_\zeta,\nu}\delta_{\mu+\nu+1,0}\delta_{M+N,0}&\cr
&+C^{-1}_{j_\zeta,\nu}\delta_{\mu+\nu-1,0}q^{(-1+\nu)}_M{}^{(-\nu)}_N
+\delta_{\mu+\nu,0}\chi^{(\nu)}_M{}^{(-\nu)}_N\Big]
\vphantom{\sum_{\xi\in\Pi}},~~~~&(5.25)\cr}$$
$$\eqalignno{
\{T_P,J_{\eta,\nu,N}\}_\kappa=&~\sum_{L\in\Bbb Z+p_{j_\eta}}\Big[
c^{(-1)}_P{}^{(-\nu)}_N{}^L_{(-\nu)}J_{\eta,\nu,L}-C^{-1}_{j_\eta,\nu}
g^{(-1)}_P{}^{(-\nu)}_N{}^L_{(-\nu+1)}J_{\eta,\nu-1,L}\Big]&\cr
&+\cdot2^{1\over 2}\kappa(t_0,t_0)\varsigma^{(-1)}_P{}^{(-1)}_N\delta_{\eta,o}
\delta_{\nu,1},&(5.26)\cr}$$
for $P,~Q\in\Bbb Z$ and $\eta,\zeta\in\Pi$, $\mu\in I_\eta$, $\nu\in
I_\zeta$, $M\in\Bbb Z+p_{j_\eta}$ and $N\in\Bbb Z+p_{j_\zeta}$.
Here,
$$\eqalignno{
c^{(m)}_M{}^{(n)}_N{}^L_{(1+m+n)}
=&~\big\langle-m\upsilon^{(m)}_M\otimes\partial\upsilon^{(n)}_N
+n\upsilon^{(n)}_N\otimes\partial\upsilon^{(m)}_M,
\upsilon^{(-m-n)}_{-L}\big\rangle,&(5.27a)\cr
f^{(m)}_M{}^{(n)}_N{}^L_{(m+n)}=
&~\big\langle\upsilon^{(m)}_M\otimes\upsilon^{(n)}_N,
\upsilon^{(1-m-n)}_{-L}\big\rangle,&(5.27b)\cr
g^{(m)}_M{}^{(n)}_N{}^L_{(m+n+2)}=&~
\big\langle\partial_\varpi\partial_\varpi\upsilon^{(m)}_M\otimes
\upsilon^{(n)}_N,\upsilon^{(-1-m-n)}_{-L}\big\rangle,&(5.27c)\cr}$$
for $m,n\in\Bbb Z/2$ and $M\in\Bbb Z+p_m$, $N\in\Bbb Z+p_n$ and
$L\in\Bbb Z+p_{m+n}$ and
$$\eqalignno{
\varsigma^{(-1)}_P{}^{(-1)}_Q=&~\big\langle\upsilon^{(-1)}_P,
D_1\upsilon^{(-1)}_Q\big\rangle,&(5.28a)\cr
\chi^{(-m)}_M{}^{(m)}_N=&~\big\langle\upsilon^{(-m)}_M,\partial_\varpi
\upsilon^{(m)}_N\big\rangle,&(5.28b)\cr
q^{(-1-m)}_M{}^{(m)}_N=&~\big\langle(R-R_\varpi)\otimes
\upsilon^{(-1-m)}_M,\upsilon^{(m)}_N\big\rangle,&(5.28c)\cr}$$
for $P,Q\in\Bbb Z$ and $m\in\Bbb Z/2$ and $M,N\in\Bbb Z+p_m$.
$o\in\Pi$, $F_{\eta,\zeta}{}^\xi$ and $N_\eta$ are defined in
sect. 2 (cf. eqs. $(2.3)$ and $(2.10)$).
$D_1$ is defined in $(4.24)$. $\partial_\varpi$ is the covariant derivative
of the connection $\varpi$: $\partial_\varpi\phi=(\partial-m\varpi)\phi$
for $\phi\in\sans K\sans N_m$. $R_\varpi$ is the meromorphic projective
connection associated to $\varpi$: $R_\varpi=\partial\varpi-(1/2)\varpi^2$.
Assume now that the poles of the meromorphic connection $\varpi$ are
simple. Then, the structure constants $f^{(m)}_M{}^{(n)}_N{}^L_{(m+n)}$,
$c^{(m)}_M{}^{(n)}_N{}^L_{(1+m+n)}$ and $g^{(m)}_M{}^{(n)}_N{}^L_{(m+n+2)}$
vanish unless $0\leq L-M-N\leq \ell,~3\ell,~5\ell$, respectively, whenever
the values of the weights, written within parenthesis, do not take the
exceptional values values $0$, $1\over 2$ for an odd theta characteristic and
$1$, while they are non zero only for finitely many values of $L-M-N$ for
the exceptional values of the weights. Similarly, $\chi^{(-m)}_M{}^{(m)}_N$,
$q^{(-1-m)}_M{}^{(m)}_N$ and $\varsigma^{(-1)}_M{}^{(-1)}_N$ vanish
unless $0\leq-M-N\leq 2\ell,~4\ell,~6\ell$, respectively, for non
exceptional values of the weights, and are non zero only for finitely many
values of $M+N$ for the exceptional values of the weights. The calculation
yielding the above formula uses $(2.1)$, $(2.2)$, $(2.3)$ and $(2.10)$ and
$(3.21)$ and is straightforward.
\par\vskip.6cm
\item{6.} {\bf The reduction of the Poisson manifold} $(\sans W,\{\cdot,
\cdot\}_\kappa)$
\vskip.4cm
\par
To obtain the classical $W$--algebras in the above framework, one has to
impose a suitable set of first class constraints on the Poisson manifold
$(\sans W,\{\cdot,\cdot\}_\kappa)$ and then fix the gauge to reduce it.
This is the subject of this section.

\proclaim Remark. In this section, $\Sigma$, $G$, $S$, $\Delta$ and $\rho$
are defined as in sect. 5.

The constraints imposed are linear. Their general form is
$$J_\Xi\approx 0,\quad \Xi\in\sans X,\eqno(6.1)$$
where $\sans X$ is some subset of $\Lie\Gau_0$
and $\approx$ denotes weak equality. Such
constraints are essentially of the same form as those used in \ref{10}
once one recalls that in the present formulation the counterpart
of the Kac--Moody current is $A+J(W)$. To implement the reduction
of $(\sans W,\{\cdot,\cdot\}_\kappa)$, one demands that the
constraints are first class. From $(5.20)$, this yields the condition
$$[\Xi,\Lambda]\in\sans X~{\rm and}~\chi(\Xi,\Lambda)=0,
\quad \Xi,\Lambda\in\sans X.\eqno(6.2)$$
One also requires that the constraint manifold is invariant under the action
of $\Conf_0$. From $(5.21)$, this yields the condition
$$\theta_u\Xi\in\sans X~{\rm and}~\chi(\dot L(u),\Xi)=0,
\quad u\in\Lie\Conf_0,\Xi\in\sans X.\eqno(6.3)$$
A maximal subspace $\sans X$ of $\Lie\Gau_0$
satisfying $(6.2)-(6.3)$ is obtained as follows. The treatment
given here follows very closely that of \ref{10}. Consider the
$2$-form $\omega\in\bigwedge^2\goth g^\vee$ defined by $\omega(x,y)=
(t_{+1},[x,y])$, $x,y\in\goth g$. The restriction of such form
to $\goth g_{-{1\over 2}}$ is non singular.
By Darboux theorem, there is a direct sum decomposition
$\goth g_{-{1\over 2}}=\goth p_{-{1\over 2}}\oplus\goth q_{-{1\over 2}}$
into subspaces of $\goth g_{-{1\over 2}}$ which are maximally isotropic and
dual to each other with respect to $\omega$. Set
$$\goth x=\goth g_{\leq -1}\oplus\goth p_{-{1\over 2}},\eqno(6.4)$$
which is a nilpotent subalgebra of $\goth g$. Then, one has
$$\sans X=\{\Xi|\Xi\in\Lie\Gau_0,~\Xi~{\rm valued~in}~
\goth x\}.\eqno(6.5)$$
This follows straightforwardly from $(5.4b)$, $(4.24)$, $(4.25)$, the
isotropy of $\goth p_{-{1\over 2}}$ with respect to $\omega$ and the
gradation of $\goth g$ by $t_0$.
{}From the theory developed in sect. 3, it is not difficult to see
that the condition of valuedness in $\goth x$ is compatible with
changes of trivializations of $L$.

The constraint manifold $\sans W_{\rm constr}$ is given in terms of the
orthogonal complement $\goth x^\perp$ of $\goth x$ with respect to
the Cartan--Killing form
$$\goth x^\perp=\goth g_{\leq 0}\oplus\ad t_{+1}\goth p_{-{1\over 2}}
\eqno(6.6)$$
and is explicitly given by
$$\sans W_{\rm constr}=\{W|W\in\sans W,~W~{\rm valued~in}~\goth x^\perp\}.
\eqno(6.7)$$
Here too, one can show that the condition of valuedness in $\goth x^\perp$
is compatible with changes of trivializations of $L$.

{}From $(4.23)$, $(4.24)$ and $(5.7)$, it follows that, for $u\in\Lie\Conf_0$
and $W\in\sans W_{\rm constr}$, $(\theta_vW)_\kappa\in\sans W_{\rm constr}$.
Thus the constraints are compatible with the action of $\Conf_0$.

{}From $(4.31)$ and $(5.13)$, it follows that, for $\Xi\in\sans X$
and $W\in\sans W_{\rm constr}$, $(\delta_\Xi W)_\kappa\in\sans W_{\rm
constr}$. The gauge symmetry, associated to the first class constraints
$(6.1)$, must be fixed. The following can be shown.

\proclaim Theorem 6.1.
For any $W\in\sans W_{\rm constr}$, there exists a unique element
$\Theta_W\in\sans X$ depending polynomially on $W$, $R$ and their derivatives
and such that
$$\ad t_{-1}(\exp\Theta_WW)_\kappa=0.\eqno(6.8)$$

{\it Proof}. The proof is quite similar to that of th. 3.12. The procedure
described by eqs. $(3.26)$ through $(3.30)$ applies also with $\Sigma$
replaced by $\Sigma\setminus\Delta$. This leads the construction of $\Theta_W$
by setting $\Omega_0=W$ and $\exp\Theta_W=\gamma_N\gamma_{N-1}\cdots\gamma_0$.
{}From $(6.4)$--$(6.5)$, it follows that $\Theta_W\in\sans X$. The argument
explained in eqs. $(3.33)$ through $(3.37)$ shows also the uniqueness of
$\Theta_W$. From $(3.5)$ and $(3.27)$, it appears that depends polynomially
on $W$, $R$ and their derivatives. Note that, unlike in the proof of
th. 3.12, $\Theta_W$ depend explicitly on $R$ since $W$ is independent from
$R$. \hfill $\square$

This theorem generalizes an analogous theorem of ref. \ref{10}. Here,
however, due account is taken of the constraints coming from the global
geometry of $\Sigma$ and $L$.

\proclaim Definition 6.1. For any $W\in\sans W_{\rm constr}$, let
$$W_c=(\exp\Theta_WW)_\kappa.\eqno(6.9)$$

{}From $(6.6)$--$(6.8)$, $W_c$ belongs to $\sans W_{\rm constr}$.
Clearly, because of the nilpotency of $\goth x$,
$W_c$ depends polynomially on $W$, $R$ and derivatives thereof.
The uniqueness of $\Theta_W$ ensures further that the map $W\rightarrow W_c$
is gauge invariant, {\it i. e.} for $\Xi\in\sans X$, and $W\in
\sans W_{\rm constr}$,
$$(\exp\Xi W)_{\kappa c}=W_c.\eqno(6.10)$$

The above suggests the following gauge fixing condition
$$W=W_c,\quad W\in\sans W_{\rm red}, \eqno(6.11)$$
defining the reduced manifold $\sans W_{\rm red}$.
$\sans W_{\rm red}$ can be characterized in terms of a set of second
class constraints. Let
$$\sans X'=\{\Xi|\Xi\in\Lie\Gau_0,~\Xi~{\rm valued~in}~
(\ker\ad t_{-1})^\perp\}.\eqno(6.12)$$
Then, $\sans W_{\rm red}$ is the submanifold of $\sans W$ determined by
$$J_\Xi\approx 0,\quad \Xi\in\sans X'\eqno(6.13)$$
and is explicitly given by
$$\sans W_{\rm red}=\{W|W\in\sans W,~W~{\rm valued~in}~\ker\ad t_{-1}\}.
\eqno(6.14)$$

It is readily verified that $(6.3)$ holds with $\sans X$ replaced
by $\sans X'$, showing that the reduced manifold is invariant under
$\Conf_0$.

$\sans W_{\rm red}$ equipped with the Dirac brackets $\{\cdot,
\cdot\}_\kappa^*$ supported on the space $\goth D(\sans W_{\rm red})$
of differential polynomials on $\sans W_{\rm red}$ defines the reduced
Poisson manifold $(\sans W_{\rm red},\{\cdot,\cdot\}_\kappa^*)$.
\par\vskip.6cm
\item{7.} {\bf The Poisson manifold} $(\sans W_{\rm red},\{\cdot,
\cdot\}_\kappa^*)$ {\bf and the classical} $W$ {\bf algebra}
\vskip.4cm
\par
The task now facing one is the computation of the Dirac brackets $\{\cdot,
\cdot\}_\kappa^*$ and the study of the properties of $\sans W_{\rm red}$.
This is the topic of this last section. In due course, a structure of
classical $W$ algebra will emerge.

\proclaim Remark. In this section, $\Sigma$, $G$, $S$, $\Delta$ and $\rho$
are defined as in sect. 5.

Any element $W\in\sans W_{\rm red}$ is completely characterized by an
ordered sequence $(w_\eta)_{\eta\in\Pi}$ with $w_\eta\in\sans K
\sans N_{j_\eta+1}$. Thus, one has the isomorphism
$$\sans W_{\rm red}\simeq\bigoplus_{\eta\in\Pi}
\sans K\sans N_{j_\eta+1},\eqno(7.1)$$
which expresses the KN content of $\sans W_{\rm red}$.
In fact, from $(6.14)$, it follows that an element $W\in\sans W$ belongs to
$\sans W_{\rm red}$ if and only if $W$ is of the form
$$W=\sum_{\eta\in\Pi}w_\eta t_{\eta,-j_\eta},\eqno(7.2)$$
where $w_\eta\in\sans K\sans N_{j_\eta+1}$.

The dual space $\sans W_{\rm red}^\vee$ of $\sans W_{\rm red}$ can be
defined as the complex vector space of ordered sequences $X=(x_\eta)_{\eta
\in\Pi}$ with $x_\eta\in\sans K\sans N_{-j_\eta}$  with the dual pairing
being given by
$$\langle X,W\rangle=\sum_{\eta\in\Pi}N_\eta\langle x_\eta,w_{\bar\eta}
\rangle\eqno(7.3)$$
see eq. $(2.10)$. Thus, one has the isomorphism
$$\sans W_{\rm red}^\vee\simeq\bigoplus_{\eta\in\Pi}\sans K\sans N_{-j_\eta}.
\eqno(7.4)$$
Since $\sans W_{\rm red}$ is a subspace of $\sans W$, it is possible to
characterize $\sans W_{\rm red}^\vee$ as the quotient of $\sans W^\vee$ by
the annihilator of $\sans W_{\rm red}$ in $\sans W^\vee$ under the non
singular dual pairing $(4.14)$. The quotient
is parametrized by assigning an element of each equivalence class.
Of course, this should be done according to a convenient criterion.
To this end, the following theorem is useful.

\proclaim Theorem 7.1.
For any $X\in\sans W_{\rm red}^\vee$ and any $V\in
\sans W_{\rm red}$, there is a unique element $E\in\sans W^\vee$ such that
$$E_{\eta,j_\eta}=x_\eta,\quad\eta\in\Pi,\eqno(7.5a)$$
$$\ad t_{-1}\big(\partial_A-\ad V\big)E=0.\eqno(7.5b)$$
Explicitly, one has
$$E=\Big[1+N\ad t_{-1}(\partial_A-\ad V)\Big]^KP_X,
\quad P_X=\sum_{\eta\in\Pi}x_\eta t_{\eta,j_\eta},\eqno(7.6)$$
where $N$ is the formal inverse of $M={1\over 2}\ad t_{-1}\ad t_{+1}$
extended by $0$ on $\ker\ad t_{+1}$ and $K\in\Bbb N,~K\geq 2j_*$,
where $j_*$ is defined in sect. 2.

{\it Proof}. The proof given here is inspired by methods developed in
ref. \ref{3}.
Let $\pi_+$ be the projector on $\ker\ad t_{+1}$ along
$\ran\ad t_{-1}$. One has
$$NM=MN=1-\pi_+,\eqno(7.7a)$$
$$[\pi_+,M]=0,\quad [\pi_+,N]=0,\eqno(7.8a-b)$$
$$[\ad t_0,M]=0,\quad [\ad t_0,N]=0.\eqno(7.8a-b)$$
Consider eq. $(7.5b)$. Next, I shall show that it can be solved
locally in any coordinate patch and give its general solution.
Using $(7.7a)$, $(7.8a)$ and $(3.21)$, one checks that $(7.5b)$ is
equivalent to
$$\Big[1-N\ad t_{-1}\big(\partial-\ad(V-Rt_{-1})\big)\Big](1-\pi_+)E=
N\ad t_{-1}\big(\partial-\ad(V-Rt_{-1})\big)\pi_+E.\eqno(7.10)$$
The operator $N\ad t_{-1}\big(\partial-\ad(V-Rt_{-1})\big)$
satisfies the relations
$$\Big[N\ad t_{-1}\big(\partial-\ad(V-Rt_{-1})\big)\Big]^K=0,
\quad K>2j_*,\eqno(7.11a)$$
$$N\ad t_{-1}\big(\partial-\ad(V-Rt_{-1})\big)=
N\ad t_{-1}\big(\partial_A-\ad V\big)+1-\pi_+.\eqno(7.11b)$$
$(7.11a)$ follows from $(7.9b)$ and the fact that $\ad t_{-1}$ lowers the
degree by $1$. $(7.11b)$ follows from $(3.21)$ and $(7.7a)$.
Recall that, for a nilpotent operator $T$, $(1-T)^{-1}$ is defined and it is
given by the series $\sum_{n=0}^\infty T^n$ containing only a finite
number of non vanishing terms. From $(7.11)$, one has then
$$\eqalignno{
E=&~\Big[1-N\ad t_{-1}\big(\partial-\ad(V-Rt_{-1})\big)\Big]^{-1}\pi_+E&\cr
=&~\sum_{n=0}^K\Big(N\ad t_{-1}\big(\partial_A-\ad V\big)+1-\pi_+\Big)^n
\pi_+E&\cr
=&~\Big(1+N\ad t_{-1}\big(\partial_A-\ad V\big)\Big)^K\pi_+E,\quad K\geq 2j_*.
&(7.12)\cr}$$
This proves that the local solution of $(7.5b)$ is completely determined
by $\pi_+E$. This suffices to show the local existence and uniqueness
of the solution of $(7.5)$. For any patch $a$, let $E_a$ be a local
solution. From $(3.1)$, it is easy to verify that
$$\ad t_{-1}\big(\partial_{A\hst1 a}-\ad V_a\big)\big(\Ad L_{ab}E_b-
E_a\big)=0.\eqno(7.13)$$
By the local uniqueness, it appears that the holomorphic $\goth g$--valued
$0$--cochain $\{E_a\}$ defines an element $E\in\sans W^\vee$ if and only if
$\pi_+(\Ad L_{ab}E_b-E_a)=0$. From $(3.3)$--$(3.4)$, this condition is
equivalent to $E_{\eta,j_\eta}\in\sans K\sans N_{j_\eta}$ for $\eta\in\Pi$.
\hfill $\square$

\proclaim Definition 7.1. For any $X\in\sans W_{\rm red}^\vee$ and any
$V\in\sans W_{\rm red}$, let $X_V$ be the element $E$ of $\sans W^\vee$
given by $(7.6)$.

For fixed $V\in\sans W_{\rm red}$, the map $X\rightarrow X_V$ defines a
linear injection of $\sans W_{\rm red}^\vee$ into $\sans W^\vee$ with the
property that
$$\langle X,W\rangle=\langle X_V,W\rangle\eqno(7.14)$$
for any $X\in\sans W_{\rm red}^\vee$ and $W\in\sans W_{\rm red}$, where
the pairing in the right hand side is the one defined by $(4.14)$. The
above relation follows from $(2.10)$, $(7.2)$ and $(7.6)$. Note that
$\langle X_V,W\rangle$ is actually independent from $V$, since only
the components $X_{V\hst1\eta,j_\eta}=x_\eta$ contribute to the result.
This expression for $\langle X,W\rangle$ is important because it can
obviously be extended to any $W\in\sans W$.

The Dirac brackets $\{\cdot,\cdot\}_\kappa^*$ are
completely defined by those of the linear functionals
$$\lambda_X(W)=\langle X,W\rangle=\lambda_{X_V}(W),
\quad W\in\sans W_{\rm red},\eqno(7.15)$$
for $X\in\sans W_{\rm red}^\vee$, where I have used $(7.14)$ and $(5.3)$
and $V$ is any element of $\sans W_{\rm red}$.

The calculation of the Dirac brackets of the $\lambda_X$ involves
the choice of a basis of $\sans X'$. Luckily, the explicit expression of the
basis elements is not necessary to carry out the calculation.

\proclaim Theorem 7.2.
For any $X,~Y\in\sans W_{\rm red}^\vee$, one has
$$\eqalignno{
\{\lambda_X,\lambda_Y\}_\kappa^*(W)=&~\langle[X_{\kappa^{-1}W},Y_0],W\rangle
+\kappa\chi(X_0,Y_0),&\cr
=&~\langle[X_0,Y_{\kappa^{-1}W}],W\rangle+\kappa\chi(X_0,Y_0),
\quad W\in\sans W_{\rm red}.&(7.16)\cr}$$

{\it Proof}. From $(5.3)$--$(5.4)$, for any $\Xi\in\Lie\Gau_0$
and $V\in\sans W_{\rm red}$, one has
$$\{J_\Xi,\lambda_{X_V}\}_\kappa(W)=\kappa\langle\Xi,(\partial_A-
\kappa^{-1}\ad W)X_V\rangle,\quad W\in\sans W.\eqno(7.17)$$
{}From $(6.12)$ and $(7.17)$, it follows that
$$\{J_\Xi,\lambda_{X_V}\}_\kappa(W)\big|_{V=\kappa^{-1}W}=0,
\quad \Xi\in\sans X',~W\in\sans W_{\rm red}.\eqno(7.18)$$
{}From this relation and the well-known formula of the Dirac brackets,
one obtains
$$\{\lambda_X,\lambda_Y\}_\kappa^*(W)=
\{\lambda_{X_V},\lambda_{Y_0}\}_\kappa(W)\big|_{V=\kappa^{-1}W}
=\{\lambda_{X_0},\lambda_{Y_V}\}_\kappa(W)\big|_{V=\kappa^{-1}W},
\quad W\in\sans W_{\rm red},\eqno(7.19)$$
In the second member, I have used the fact that the Dirac bracket is
independent form the extension $\lambda_{Y_V}$ of $\lambda_Y$ to $\sans W$
used to set $V=0$ and analogously in the third member.
The cocycle term is
$$\langle X_{\kappa^{-1}W},\partial_A Y_0\rangle=
-\langle Y_{\kappa^{-1}W},\partial_A X_0\rangle
=\langle X_0,\partial_A Y_0\rangle,\quad W\in\sans W_{\rm red},\eqno(7.20)$$
since, by $(7.5b)$, $\partial_AX_0,~\partial_AY_0\in\sans W_{\rm red}$ and
$\langle X_V,W\rangle$ is independent from $V\in\sans W_{\rm red}$
for any $W\in\sans W_{\rm red}$. \hfill $\square$

The first term in the right hand side of $(7.16)$ is a differential
polynomial in the $x_\eta$, $y_\eta$ and $w_\eta$ and is computable
in principle using $(7.16)$. The second term, proportional to $\kappa$, is
the anomaly. It can be calculated explicitly. The result is
$$\chi(X_0,Y_0)=\sum_{\eta\in\Pi}N_\eta
\bigg[\prod_{m\in I_\eta, m\geq-j_\eta+1}{2\over C^{-1}{}_{j_\eta,m}}\bigg]
\langle x_\eta,D_{j_\eta}y_{\bar\eta}\rangle,\eqno(7.21)$$
$$\eqalignno{D_0=&~\partial,&\cr
D_{1\over2}=&~\partial^2+(1/2)R,&\cr
D_1=&~\partial^3+2R\partial+(\partial R),&\cr
D_{3\over 2}=&~\partial^4+5R\partial^2+5(\partial R)\partial
+(3/2)\big(\partial^2R+(3/2)R^2\big),&\cr
D_2=&~\partial^5+10 R\partial^3+15(\partial R)\partial^2+
[9(\partial^2R)+16R^2]\partial+2[(\partial^3R)+8R(\partial R)],&\cr
&{\rm etc.}.&(7.22)\cr}$$
The $D_j$ are the well-known Bol operators \ref{21}.

There are other relevant Dirac brackets. Consider the
energy-momentum tensor $T$. For any $u\in\Lie\Conf_0$, the
restriction of $T_u$ to $\sans W_{\rm red}$, which will be denoted
by the same symbol, is given by $(5.9b)$ with $W\in\sans W_{\rm red}$.
Explicitly,
$$T_u=2^{-{1\over 2}}(t_0,t_0)\langle u,w_o\rangle
+(1/2\kappa)\sum_{\eta\in\Pi,j_\eta=0}N_\eta
\langle u,w_\eta{}^{\otimes 2}\rangle,\eqno(7.23)$$
where $o\in\Pi$ is defined in sect. 2.
As appears, $T_u\in\goth D(\sans W_{\rm red})$. Note that only the
components $w_\eta$ with $\eta\in\Pi$ and $j_\eta=0$, which correspond to
$\ker\ad t_{-1}\cap\goth c_{\goth s}$, contribute to the term
quadratic in $W$ \footnote{${}^2$}{In ref. \ref{6}, such quadratic
contribution was overlooked.}.
{}From $(7.16)$, one has
$$\{T_u,T_v\}_\kappa^*=T_{[u,v]}+12\kappa(t_0,t_0)\sigma(u,v)
\eqno(7.24)$$
for any $u,~v\in\Lie\Conf_0$, which is to be compared with
$(5.17)$. Using $(7.16)$, one also finds
$$\{T_u,\lambda_X\}^*_\kappa=\lambda_{\theta_vX}+\kappa\chi(\dot L(u),X_0)
\eqno(7.25)$$
for any $u\in\Lie\Conf_0$ and any $X\in\sans W_{\rm red}^\vee$,
where $\theta_u X=(\theta_u x_\eta)_{\eta\in\Pi}$ is given by $(4.11)$ with
$\phi=x_\eta$ and $h=-j_\eta$. The calculations involved in deducing
$(7.24)$ and $(7.25)$ are straightforward.

Let us discuss briefly the results just obtained. $(7.16)$ defines a Dirac
bracket $W$ algebra in the so-called lowest weight gauge. In fact,
analogous expressions have been worked out in the literature following
closely related techniques (see refs. \ref{9--13}).
The $W$ algebra proper is obtained by letting $x_\eta$ and $y_\eta$
in $(7.16)$ be elements of the KN basis of $\sans K\sans N_{-j_\eta}$.
The form of the anomaly was first found in \ref{29} in a different
approach where however the deep relation with the theory of $SL(2,\Bbb C)$
embeddings into simple Lie groups was not apparent.
{}From $(7.23)$, it follows that the $T_u$ form a Dirac bracket
Virasoro algebra of classical
central charge $12\kappa(t_0,t_0)$. From $(7.24)$, it also appears that
the functions $\lambda_X$ with $x_o=0$ are primary with respect to the Virasoro
algebra. All the above properties have a counterpart in the standard algebraic
formulation to $W$ algebras \ref{9--13}.

One may consider the $W$ algebra obtained above in terms of modes.
For any $\eta\in\Pi$ and any $M\in\Bbb Z+p_{j_\eta}$, let $X_{\eta,M}$
be the element of $\sans W_{\rm red}^\vee$ defined by the ordered sequence
$(\delta_{\eta,\zeta}\upsilon^{(-j_\eta)}_N)_{\zeta\in\Pi}$. Set
$$j_{\eta,M}=\lambda_{X_{\eta,M}}.\eqno(7.26)$$
{}From $(7.16)$, by means of a straightforward calculation, one finds,
to order $O(\kappa^0)$,
$$\eqalignno{\{j_{\eta,M},j_{\zeta,N}\}_\kappa^*=&~
\kappa N_\eta\delta_{\eta,\bar\zeta}
\bigg[\prod_{m\in I_\eta, m\geq-j_\eta+1}{2\over C^{-1}{}_{j_\eta,m}}\bigg]
\varsigma^{(-j_\eta)}_M{}^{(-j_\eta)}_N&\cr
&+\sum_{\xi\in\Pi}\sum_{L\in\Bbb Z+p_{j_\xi}}
F_{\eta,\zeta}{}^\xi h^{(j_\eta)}_M{}^{(j_\zeta)}_N{}^L_{(1+j_\xi)}j_{\xi,L}
+O(\kappa^{-1}),&(7.27)\cr}$$
for $\eta,\zeta\in\Pi$, $M\in\Bbb Z+p_{j_\eta}$ and $N\in\Bbb Z+p_{j_\zeta}$.
Here,
$$\eqalignno{
h^{(j_\eta)}_M{}^{(j_\zeta)}_N{}^L_{(1+j_\xi)}=&~
\bigg\langle\sum_{m\in I_\eta,n\in I_\zeta,m+n=j_\xi}(j_\eta,m;j_\zeta,n|
j_\xi,j_\xi)X_{\eta,M\hst1 0\hst1 \eta,m}X_{\zeta,N\hst1 0\hst1 \zeta,n}
,\upsilon^{(1+j_\xi)}_{-L}\bigg\rangle,&\cr
\varsigma^{(-j_\eta)}_M{}^{(-j_\eta)}_N=&~\big\langle\upsilon^{(-j_\eta)}_M,
D_{j_\eta}\upsilon^{(-j_\eta)}_N\big\rangle,&\cr
X_{\eta,M\hst1 0}=&~F_{0,\eta,j_\eta}\big(\upsilon^{(-j_\eta)}_M\big),
\quad{\rm etc.}&(7.28)\cr}$$
(cf. eq. $(3.12)$) and $F_{\eta,\zeta}{}^\xi$ and $N_\eta$ are defined in
sect. 2 (cf. eqs. $(2.3)$ and $(2.10)$). It can also be seen that
$h^{(j_\eta)}_M{}^{(j_\zeta)}_N{}^L_{(1+j_\xi)}$ and
$\varsigma^{(-j_\eta)}_M{}^{(-j_\eta)}_N$ vanish unless
$j_\eta+j_\zeta-2j_\xi\leq L-M-N\leq\big[2(j_\eta+j_\zeta-j_\xi)
+1\big]\ell+2j_\xi-j_\eta-j_\zeta$ and $-2(2j_\eta+1)\ell\leq M+N\leq 0$,
respectively, if the weights involved are non exceptional. The expression
of $X_{\eta,M\hst1 0}$ follows easily from noting that the equation
$\ad t_{-1}\partial_A X_{\eta,M\hst1 0}=0$ obeyed by $X_{\eta,M\hst1 0}$
is equivalent to $(3.5)$ with $h=0$, $\mu=j_\eta$ and $\phi
=\upsilon^{(-j_\eta)}_N$.
\vskip.6cm
\par\noindent
{\bf Acknowledgements.} I wish to voice my gratitude to M. Matone,
E. Aldrovandi and expecially R. Stora for helpful discussions.
\vskip.6cm
\centerline{\bf REFERENCES}
\def\ref#1{\lbrack #1\rbrack}
\def\NP#1{Nucl.~Phys.~{\bf #1}}
\def\PL#1{Phys.~Lett.~{\bf #1}}

\def\CMP#1{Commun.~Math.~Phys.~{\bf #1}}
\def\PR#1{Phys.~Rev.~{\bf #1}}

\def\LMP#1{Lett.~Math.~Phys.~{\bf #1}}

\def\IJMP#1{Int.~J.~Mod.~Phys.~{\bf #1}}
\def\PREP#1{Phys.~Rep.~{\bf #1}}

\def\AP#1{Ann.~Phys.~{\bf #1}}
\def\CQG#1{Class.~Quantum~Grav.~{\bf #1}}
\vskip.4cm
\par\noindent

\item{\ref{1}}
P. Bouwknegt and K. Schoutens, \PREP{223} (1993) 183 and references therein.

\item{\ref{2}}
R. Zucchini, \CQG{10} (1993) 253.

\item{\ref{3}}
J. De Boer and J. Goeree, \NP{B401} (1993) 369.

\item{\ref{4}}
E. Aldrovandi and L. Bonora, LANL hep-th/9303064.

\item{\ref{5}}
E. Aldrovandi and G. Falqui, LANL hep-th/9312093.

\item{\ref{6}}
R. Zucchini, LANL hep-th/9310061, to appear in \PL{B}.

\item{\ref{7}}
I. M. Krichever and S. P. Novikov, Funktz. Analiz.
Prilozhen. {\bf 21} n. 2 (1987) 46, Funktz. Analiz. Prilozhen. {\bf 21}
n. 4 (1987) 47, Funktz. Analiz. Prilozhen. {\bf 23} n. 1 (1989) 19.

\item{\ref{8}}
V. G. Drinfeld and V. V. Sokolov, J. Sov. Math. {\bf 30} (1985) 1975.

\item{\ref{9}}
J. Balog, L. Feh\'er, L. O'Raifeartaigh,
P. Forg\'acs and A. Wipf, \AP{203} (1990) 76.

\item{\ref{10}}
L. Feh\'er, L. O'Raifeartaigh, P. Ruelle, I. Tsutsui and A. Wipf, \PR{222}
no. 1 (1992) 1.

\item{\ref{11}}
F. A. Bais, T. Tjin and P. van Driel, \NP{B357} (1992) 632.

\item{\ref{12}}
T. Tjin, {\it Finite and Infinite W--algebras and their Applications},
Doctoral Thesis, LANL hep-th/9308146.

\item{\ref{13}}
A. Sevrin and W. Troost, \PL{B315} (1993) 304.

\item{\ref{14}}
J.-L. Gervais, L. O'Raifeartaigh, A. V. Razumov and M. V. Saveliev,
\PL{B301} (1993) 41.

\item{\ref{15}}
A. N. Leznov and M. V. Saveliev, \PL{B79} (1978) 294,
\LMP{3} (1979) 207, \CMP{74} (1980) 111, \LMP{6} (1982) 505, \CMP{89}
(1983) 59.

\item{\ref{16}}
A. Bilal and J.-L. Gervais, \PL{B206} (1988) 412, \NP{B314} (1989) 646.

\item{\ref{17}}
E. B. Dynkin, Am. Math. Soc. Transl. {\bf 6} (2) (1967) 111.

\item{\ref{18}}
N. Jacobson, {\it Lie Algebras}, Wiley--Interscience, New York, 1962.

\item{\ref{19}}
D. M. Brink and G. R. Satchler, {\it Angular Momentum}, Oxford University
Press 1962.

\item{\ref{20}}
R. Zucchini, LANL hep-th/9307015.

\item{\ref{21}}
F. Gieres, \IJMP{A8} (1993) 1.

\item{\ref{22}}
R. Gunning, {\it Lectures on Riemann Surfaces}, Princeton University
Press 1966.

\item{\ref{23}}
R. Gunning, {\it Lectures on Vector Bundles on Riemann Surfaces}, Princeton
University Press 1967.

\item{\ref{24}}
M. Schlichenmaier, Lett. Math. Phys. {\bf 19} (1990) 151, Lett. Math. Phys.
{\bf 19} (1990) 327, Lett. Math. Phys. {\bf 20} (1990) 33.

\item{\ref{25}}
R. Dick, Lett. Math. Phys. {\bf 18} (1989) 255; Fortschr. Phys. {\bf 40}
(1992) 519.

\item{\ref{26}}
L. Bonora, A. Lugo, M. Matone and J. Russo, \CMP{123} (1989) 329.

\item{\ref{27}}
S. Klimek and A. Lesniewski, \CMP{125} (1989) 597.

\item{\ref{28}}
V. A. Sadov, \CMP{136} (1991) 585.

\item{\ref{29}}
M. Matone, {\it Conformal Field Theories in Higher Genus},
Doctoral Thesis, SISSA--ISAS, Trieste, Italy.

\bye